\documentclass[12pt]{iopart}
\usepackage{iopams}
\usepackage{epsfig}

\begin{document}

\tolerance = 10000


\title{Dynamical Functions of a 1D Correlated Quantum Liquid}
\author{J. M. P. Carmelo$^1$, D. Bozi$^2$ and K. Penc$^3$} 
\address{$^1$GCEP-Center of Physics, U. Minho, Campus Gualtar,
P-4710-057 Braga, Portugal\\
$^2$Centro de F\'{\i}sica de Materiales, 
Centro Mixto CSIC-UPV/EHU,
E-20018 San Sebastian, Spain\\
$^3$Res. Inst. for Solid State Physics
and Optics, H-1525 Budapest, P.O.B. 49, Hungary\\
E-mail: carmelo@fisica.uminho.pt}

\begin{abstract}
The dynamical correlation functions in the one-dimensional electronic systems show 
power-law behavior at low energies and momenta close to integer multiples 
of the charge and spin {\it Fermi momenta}. These systems are usually
referred to as Tomonaga-Luttinger liquids. However, near well-dfined
lines of the $(k,\omega)$ plane the power-law behaviour extends beyond the
low-energy cases mentioned above, and also appears at higher energies 
leading to singular features in the photoemission spectra and other 
dynamical correlation functions. The general spectral-function
expressions derived in this paper were used in recent theoretical
studies of the finite-energy singular features in photoemission of the organic compound
tetrathiafulvalene-tetracyanoquinodimethane (TTF-TCNQ) metallic phase. They are based on a so called
pseudofermion dynamical theory (PDT), which allows us to systematically enumerate and
describe the excitations in the Hubbard model starting from the Bethe-Ansatz, as well as to
calculate the charge and spin objects phase shifts appearing 
as exponents of the power laws. In particular, we concentrate
on the spin-density $m\rightarrow 0$ limit and in effects of the vicinity of 
the singular border lines, as well as close to half filling. 
Our studies take into account spectral contributions from 
types of microscopic processes that do not occur for finite 
values of the spin density. In addition, the specific processes 
involved in the spectral features of TTF-TCNQ are studied. 
Our results are useful for the further understanding of the 
unusual spectral properties observed in 
low-dimensional organic metals and also provide expressions 
for the one- and two-atom spectral functions of a correlated 
quantum system of ultracold fermionic atoms in a 1D optical 
lattice with on-site two-atom repulsion. 
\end{abstract}

\pacs{71.00}

\maketitle
\section{INTRODUCTION}
\label{SecI}

The low-energy physics of correlated one-dimensional (1D) problems 
has some universal properties described by the Tomonaga-Luttinger 
liquid (TLL) \cite{Voit}. In turn, a pseudofermion dynamical theory (PDT)
beyond the TLL \cite{V-1,V} was recently used to study the
finite-energy singular features in 
photoemission of the organic compound tetrathiafulvalene-tetracyanoquinodimethane 
(TTF-TCNQ) metallic phase \cite{TTF}. While the PDT was originally introduced
for the 1D Hubbard model, more recently other methods 
for the study of finite-energy spectral and dynamical functions of
1D correlated systems were
introduced \cite{Affleck,Glazman}. Both the finite-energy spectral-weight
distributions studied by the PDT for the 1D Hubbard model and the methods of
Refs. \cite{Affleck,Glazman} for other 1D correlated problems include 
power-law singularities near
well-defined branch lines with exponents depending on the interaction
strength and the excitation momentum. In the limit of low energy
the PDT correlation- and spectral-function expressions recover
the usual low-energy TLL results \cite{LE} 

Besides a renewed interest in the unusual spectral and dynamical
properties of quasi-1D organic compounds \cite{Zwick,spectral0},
recently there has been an increasing interest in those of 
interacting ultracold fermionic atoms in 1D
optical lattices \cite{Zoller}. Quantum effects are strongest at low dimensionality 
leading to unusual phenomena such as charge-spin separation at all energies 
\cite{Zwick,spectral0}. Thus, the further understanding of the microscopic mechanisms 
behind the unusual spectral properties observed in low-dimensional correlated systems 
and materials is a topic of high scientific interest. 

The 1D Hubbard model is one of the few realistic models for correlated electrons in a
discrete lattice for which one can exactly calculate all the energy eigenstates and their
energies \cite{Lieb,Takahashi,HL}. It includes a first-neighbor transfer-integral $t$, for
electron hopping along the chain, and an effective on-site Coulomb repulsion $U$. For
finite-energy values the metallic phase of this model goes beyond the low energy behavior 
described by the usual TLL 
\cite{Voit} and thus the study of spectral functions is a very involved 
many-electron problem. Fortunately, the construction of a pseudofermion description
by means of a unitary transformation which slightly shifts the discrete 
momentum values of the corresponding pseudoparticles
of Refs. \cite{I,II-04} leads to the PDT, whose energy spectrum has no residual-interaction
energy terms. Therefore, such a description is suitable for the derivation of explicit expressions for
these functions \cite{V-1,V,LE}. 

The PDT is a generalization for finite values of $U$ of the scheme introduced in Refs.
\cite{Penc,Karlo} for large $U$ values. It profits from the use of a description of
the exact energy eigenstates in terms of occupancy configurations of several
branches of pseudofermions. The ground state has finite occupancy of
charge $c$ and spin $s1$ pseudofermions only.
Under the ground-state
- excited-energy-eigenstate transitions, the 
pseudofermions and pseudofermion holes undergo
elementary scattering events with the pseudofermions
and pseudofermion holes created in these transitions. This leads to 
excited-state, interaction, density, and two-momenta dependent
two-pseudofermion phase shifts. The point is
that the one- and two-electron spectral functions can
be expressed in terms of pseudofermion determinants which 
are a functional of such phase shifts. Use of the PDT reveals that for finite
values of $U/t$ all singular one-electron spectral features of the model are of power-low
type, controlled by negative exponents. Furthermore, the PDT line shapes 
associated with such exponents were found to correspond to the unusual 
charge and spin spectral features observed by photoemission experiments 
for the whole finite-energy band width in quasi-1D organic metals \cite{spectral}.
(The use of the Dynamical Density Matrix Renormalization Group method 
\cite{Eric} leads to results consistent with those obtained by the PDT.) Furthermore, 
when combined with the Renormalization Group, the use of the PDT reveals that 
a system of weakly coupled Hubbard chains is suitable for the successful
description of the phase diagram observed in quasi-1D doped Mott-Hubbard insulators
\cite{super}. 

The low-energy physics of the model corresponds to the universal TLL behaviour
and was studied by different techniques, such as conformal-field theory  
\cite{CFT} and bosonization \cite{Bozo}. There are many investigations where the
low-energy conformal invariance was combined with the model exact Bethe-ansatz solution
in the study of the asymptotics of correlation functions and related quantities
\cite{Voit,Bozo}. As mentioned above, the studies of Ref. \cite{LE} 
confirm that in the limit of low energy the general finite-energy 
spectral and correlation function expressions provided by the PDT in Ref. \cite{V-1} 
recover the correct behaviour given by conformal-field theory. 

The main goal of this paper is to provide the details of
an extension of the PDT introduced in Refs. \cite{V-1,V,LE} 
to initial ground states with spin density $m\rightarrow 0$ which
was used in recent theoretical studies of the finite-energy singular features in 
photoemission of the 
TTF-TCNQ metallic phase \cite{TTF}. The point
is that such an extension involves spectral contributions 
that do not occur for finite 
values of the spin density and hence are not taken into
account by the expressions of Refs. \cite{V-1,V,LE}.
In order to introduce the generalized spectral-function expressions
used in the recent studies of the TTF-TCNQ spectral
features presented in short form in Ref. \cite{TTF} 
we present here such an extension of the PDT.

The unusual spectral properties are mainly determined by the
occupancy configurations in the excited states 
of the two pseudofermion branches 
that have finite occupancies for the ground state. 
Those are the charge $c0$ pseudofermion and spin 
$s1$ pseudofermion branches. The ground-state corresponds to 
charge $c0$ pseudofermion (and spin $s1$ pseudofermion) finite 
occupancy for canonical-momentum values in the range 
$\vert\bar{q}\vert<q^0_{Fc0}=2k_F$ (and $\vert\bar{q}\vert<q^0_{Fs1}=
k_{F\downarrow}$) and charge $c0$ pseudofermion-hole (and spin 
$s1$ pseudofermion-hole) finite occupancy for canonical-momentum 
values in the range $2k_F<\vert\bar{q}\vert<q^0_{c0}=\pi$ (and 
$k_{F\downarrow}<\vert\bar{q}\vert<q^0_{s1}=k_{F\uparrow}$).
The values of the four deviations $\Delta {\bar{q}}_{Fc0,\,\pm 1}$ and
$\Delta {\bar{q}}_{Fs1,\,\pm 1}$ in the two charge 
$\pm q^0_{Fc0}=\pm 2k_F$ and two spin $\pm q^0_{Fs1}=
\pm k_{F\downarrow}$ {\it Fermi} points under each ground-state
- excited-energy-eigenstate transition play a major role
in the PDT. 

We denote such deviations by $\Delta {\bar{q}}_{F\alpha\nu ,\,\iota}$ 
where $\alpha\nu= c0,\,s1$ and $\iota =\pm 1$. The spectral-weight
distributions are controlled by the values of the following associated
four parameters,
\begin{equation}
2\Delta_{\alpha\nu}^{\iota} = \left({\Delta
{\bar{q}}_{F\alpha\nu ,\,\iota}\over [2\pi/L]}\right)^2 \, ;
\hspace{0.25cm} \alpha\nu= c0,\,s1\, ;
\hspace{0.25cm} \iota =\pm 1 \, ,
\label{Delta}
\end{equation}
where $L>>1$ is the 1D lattice length within the use of periodic
boundary conditions. According to the PDT, the contributions from 
ground-state transitions to subspaces spanned by sets of excited 
energy eigenstates with the same values for the two charge 
parameters $2\Delta_{c0}^{\pm 1}$ and two spin parameters 
$2\Delta_{s1}^{\pm 1}$ fully determine the momentum $k$ and 
energy $\omega$ dependence of the general finite-energy
spectral functions in the small $(k,\omega)$-plane region 
associated with the energy and momentum spectrum
of these excited states \cite{V-1,V}. 

The general PDT finite-energy spectral-function expressions given in Ref. \cite{V-1}
refer to the metallic phase for initial ground states 
with spin densities $m$ in the range $0<m<n$, where
$n$ is the electronic density such that $0<n<1$. Actually, the expressions
found in Ref. \cite{V-1} refer to values of $n$ such that the {\it Fermi}-point charge 
velocity $v_{c0}$ is larger than the {\it Fermi}-point spin velocity $v_{s1}$.
For finite values of $U/t$, this excludes densities $n$ in the vicinity
of half filling. One of our goals is to extend the studies of that reference
to electronic densities such that $v_{s1}>v_{c0}$.

The general expressions of the exponents that control the singular 
features of the above finite-energy spectral function expressions provide 
the correct zero spin-density values in the limit $m\rightarrow 0$. 
Moreover, as given in Eq. (55) of Ref. \cite{V-1}, for excitations such that 
$2\Delta_{\alpha\nu}^{\iota}\neq 0$ for the two $\alpha\nu=c0$ charge and two 
$\alpha\nu=s1$ spin parameters, the corresponding four pseudofermion relative weights
have the asymptotic expression provided in that equation. It follows
that for zero-spin excitations such that the four parameters
$2\Delta_{\alpha\nu}^{\iota}$ where $\alpha\nu=c0,s1$ and $\iota=\pm 1$
are finite, the general convolution 
function and corresponding pre-factor function $F_0 (z)$ given 
in Eqs. (61) and (62) of Ref. \cite{V-1}, respectively, provide the correct zero
spin-density contributions to the spectral-function expressions in the limit 
of zero spin density. The point is that for densities in the ranges $0<n<1$ and 
$0<m<n$ all ground-state - excited-energy-eigenstate transitions 
lead to finite values for these four parameters. 

While some $m\rightarrow 0$ one- and two-electron excitations also
lead to finite values for the two $\alpha\nu=c0$ charge 
and two $\alpha\nu=s1$ spin parameters 
$2\Delta_{\alpha\nu}^{\iota}$, there are also $m\rightarrow 0$ excitations for 
which one (or both) spin parameter(s) $2\Delta_{s1}^{\pm 1}$ vanishes (or 
vanish). In that case the corresponding
$s1,\iota$ pseudofermion relative weights have not the
asymptotic expression provided in Eq. (55) of Ref. \cite{V-1}.
Nonetheless, we find in this paper that the contributions to the 
spectral-function expressions from ground-state transitions 
to the excited energy eigenstates which span such excitations
lead to convolution functions of the same general form as that
provided in Eq. (61) of Ref. \cite{V-1}. The only difference is that 
the pre-factor $F_0 (z)$ of such convolution functions given in 
Eq. (62) of that reference is replaced by another suitable function
derived in this paper and given in Appendix A.

Since these convolution functions fully control all the different 
spectral-function contributions given in Eqs. (66), (68), and (70) 
of Ref. \cite{V-1}, the derivation of all $m=0$ contributions requires 
the use of such suitable pre-factors, which were taken into account 
in the studies of the TTF-TCNQ spectral features of Ref. \cite{TTF}
and we calculate in this 
paper. Interestingly, the different expressions found here for the pre-factor
$F_0 (z)$ are such that the corresponding pre-factors of the space 
and time asymptotic expressions of correlation functions \cite{LE}
are continuous functions of $m$ as $m\rightarrow 0$. 

The extension of the expressions for the 
one- and two-electron spectral-weight distributions introduced 
in Ref. \cite{V-1} for the 1D Hubbard model to 
the regime where $m\rightarrow 0$ involves taking into
account contributions from
transitions to subspaces spanned by excited states such that one, two,
three, or even four out of the four parameters $2\Delta_{\alpha\nu}^{\iota}$
where $\alpha\nu = c0,s1$ and $\iota =\pm 1$
vanish. Note that according to Eq. (\ref{Delta}), the value of the 
$\alpha\nu ,\iota$ pseudofermion canonical-momentum {\it Fermi} points 
associated with $2\Delta_{\alpha\nu}^{\iota}=0$ values remains 
unchanged under the corresponding ground-state - excited-state transitions. 
Such contributions do not exist for the $m>0$ regime addressed in 
Ref. \cite{V-1} but must be taken into account in the quantitative study 
of the spectral-weight distributions of the correlated metal at zero spin density. 
In this paper we also extend the one- and two-electron 
spectral-weight distributions of Ref. \cite{V-1} to all electronic densities of the metallic
phase. Although we do not consider the half-filling Mott-Hubbard 
insulator such that $v_{c0}=0$, our expressions refer to all electronic 
densities of the metallic phase and thus also for those in the vicinity of $1$. 

Another problem solved in this paper is the derivation of explicit expressions 
for the spectral functions in the vicinity of the singular border lines, which
again were used in the recent studies of the TTF-TCNQ spectral features 
of Ref. \cite{TTF}. For that one-electron problem such border lines correspond 
to processes for which the extra charge and spin objects created upon 
removal or addition of the electron have exactly the same velocity. Such
an equality of the charge and spin velocities occurs at well-defined lines in 
the $(k,\omega)$ plane and leads to power-law 
singular behavior along such a border lines. 
In contrast to the power-law branch-line singular features
studied in Ref. \cite{V-1}, which are controlled by momentum, $U/t$, and density
dependent negative exponents, the border-line power-law singular
features are controlled by an universal exponent given by
$-1/2$, as given in Eq. (2) of Ref. \cite{TTF}. Further details on the
processes involved in the applications to TTF-TCNQ are
also reported.

The paper is organized as follows: In Sec. \ref{SecII} we
introduce the model and provide basic information about the
pseudofermion description needed for our studies.
In Sec. \ref{SecIII} the general expressions required
for the study of the finite-energy spectral-weight distributions 
of the metallic phase for initial $m\rightarrow 0$ ground 
states are calculated. That includes derivation
of explicit expressions of the spectral functions in 
the vicinity of the singular border lines for density 
ranges $0<n<1$ and $0\leq m<n$. 
In Sec. \ref{SecIV} we use the expressions obtained
in the previous section and in Ref. \cite{V-1}
to provide further details on the processes that
contribute to the unusual photoemission spectrum 
of TTF-TCNQ. Finally, the concluding remarks 
are presented in Sec. \ref{SecV}.

\section{THE PROBLEM AND THE PSEUDOFERMION DESCRIPTION} \label{SecII}

Our study focus on finite-$\omega$ $\cal{N}$-electron spectral-weight distributions 
of the following general form,
\begin{equation}
B_{{\cal{N}}}^{l} (k,\,\omega) = \sum_{f}\, \vert\langle f\vert\,
{\hat{O}}_{{\cal{N}}}^{l} (k) \vert GS\rangle\vert^2\,\delta\Bigl(
\omega - l[E_f - E_{GS}]\Bigr) \, ; \hspace{0.25cm} l\omega > 0 \, , 
\label{ABON}
\end{equation}
where $l = \pm 1$. We focus our attention on the cases of more physical interest
which correspond to ${\cal{N}}=1,\,2$. In the above expression
the general $\cal{N}$-electron operators
${\hat{O}}_{{\cal{N}}}^{+1} (k)\equiv {\hat{O}}_{{\cal{N}}}^{\dag}
(k)$ and ${\hat{O}}_{{\cal{N}}}^{-1} (k) \equiv
{\hat{O}}_{{\cal{N}}} (k)$ carry momentum $k$, the $f$ summation
runs over the excited energy eigenstates, the energy $E_f$
corresponds to these states, and $E_{GS}$ is the initial ground-state
energy. The local operator ${\hat{O}}_{{\cal{N}},\,j}^{+1}\equiv
{\hat{O}}_{{\cal{N}},\,j}^{\dag}$ or
${\hat{O}}_{{\cal{N}},\,j}^{-1}\equiv {\hat{O}}_{{\cal{N}},\,j}$
is related to the corresponding momentum-representation operator
${\hat{O}}_{{\cal{N}}}^{l} (k)$ of Eq. (\ref{ABON}) by a Fourier
transform. As in Ref. \cite{V-1}, we use in expression
(\ref{ABON}) a momentum extended scheme such that $k\in
(-\infty,\,+\infty)$.

We consider weight distributions (\ref{ABON}) that refer to
the Hubbard model in a 1D lattice with periodic boundary conditions 
and units such that the Planck constant and electronic lattice constant are one, 
\begin{equation}
\hat{H}=-t\sum_{j,\,\sigma}[c_{j,\,\sigma}^{\dag} c_{j+1,\,\sigma}
+ h. c.]+U\sum_{j}\hat{n}_{j,\uparrow}\hat{n}_{j,\downarrow} \, .
\label{H}
\end{equation}
Here $c_{j,\,\sigma}^{\dagger}$ (and $c_{j,\,\sigma}$) creates (and
annihilates) one spin-projection $\sigma =\uparrow ,\downarrow$ electron
at site $j=1,2,...,N_a$ and $\hat{n}_{j,\,\sigma}=c_{j,\,\sigma}^{\dagger}\,c_{j,\,\sigma}$.
Let $N=N_{\uparrow}+N_{\downarrow}$ be the electronic number, $N_a$ the number 
of lattice sites, and $n_{\sigma}=N_{\sigma}/L=N_{\sigma}/N_a$. $N_a$ is assumed 
to be even and very large. 
The electronic densities $n=n_{\uparrow }+n_{\downarrow}$ 
and spin densities $m=n_{\uparrow}-n_{\downarrow}$ are in the ranges 
$0<n< 1$ and $0\leq m<n$, respectively. Except for corrections of order
of $1/L$, the Fermi momenta are given by $k_F=\pi n/2$ and
$k_{F\sigma}=\pi n_{\sigma}$. 

The concept of a rotated electron plays a key role in the
pseudofermion description. Concerning its relation to the holons,
spinons, and $c0$ pseudoparticles whose occupancy configurations
describe the energy eigenstates of the model (\ref{H}), see Ref.
\cite{I}. Our studies do not involve directly the holons and
spinons as defined in that reference. The charge $c\nu$ pseudofermions (and 
spin $s\nu$ pseudofermions) such that $\nu=1,2,3,...$ are $2\nu$-holon  
(and $2\nu$-spinon) composite quantum objects whose discrete momentum 
values are slightly shifted relative to those of the corresponding $c\nu$ 
pseudoparticles (and spin $s\nu$ pseudoparticles) studied in Refs. \cite{I,II-04}. 
Such momentum shifts cancel exactly the residual-interaction terms
of the pseudoparticle energy spectrum. Otherwise, pseudoparticles
and pseudofermions have the same properties. 

According to the PDT of 
Refs. \cite{V-1,LE}, the charge $c0$ pseudofermions and spin $s1$ 
pseudofermions play the major role in the spectral properties.
The holons (and spinons) which are not part of $2\nu$-holon 
composite $c\nu$ pseudofermions (and $2\nu$-spinon composite 
$s\nu$ pseudofermions) are the Yang holons
(and HL spinons). Those are invariant under the 
electron - rotated-electron unitary transformation and hence
have a non-interacting character and do not contribute
to the matrix elements between energy eigenstates 
of the spectral-weight distributions studied in this paper.
We denote the numbers of $\alpha\nu$
pseudofermions and $\alpha\nu$ pseudofermion holes by
$N_{\alpha\nu}$ and $N^h_{\alpha\nu}$, respectively, where $\alpha
=c,\,s$ and $\nu =0,1,2,...$ for $\alpha =c$ and $\nu =1,2,...$
for $\alpha =s$. (The value of $N^h_{\alpha\nu}$ is given in Eqs.
(B7) and (B8) of Ref. \cite{I}.) 

Alike in Ref. \cite{V-1}, we use in this paper the notation $\alpha\nu\neq c0,\,s1$
branches, which refers to all $\alpha\nu$ branches except the $c0$
and $s1$ branches. Moreover, the summations (and products)
$\sum_{\alpha\nu}$, $\sum_{\alpha\nu =c0,\,s1}$, and
$\sum_{\alpha\nu\neq c0,\,s1}$ (and $\prod_{\alpha\nu}$,
$\prod_{\alpha\nu =c0,\,s1}$, and $\prod_{\alpha\nu\neq c0,\,s1}$)
run over all $\alpha\nu$ branches with finite $\alpha\nu$
pseudofermion occupancy in the corresponding state or subspace,
the $c0$ and $s1$ branches only, and all $\alpha\nu$ branches with
finite $\alpha\nu$ pseudofermion occupancy in the corresponding
state or subspace except the $c0$ and $s1$ branches, respectively.

The pseudofermion description refers to a Hilbert subspace called
the {\it pseudofermion subspace} (PS) in Ref. \cite{V-1}, in
which the $\cal{N}$-electron excitations
${\hat{O}}_{{\cal{N}}}^{l} (k)\vert GS\rangle$ are contained. The
PS is spanned by the initial ground state and the excited energy
eigenstates originated from it by creation, annihilation, and
particle-hole processes involving the generation of a finite
number of active pseudofermion scattering centers, Yang
holons, and HL spinons plus a vanishing or small density of
low-energy and small-momentum $\alpha\nu =c0,\,s1$ pseudofermion
particle-hole processes. It is convenient to classify these processes
into three types, called processes (A), (B), and (C), as further 
discussed in the ensuing section. 

The $\alpha\nu$-pseudofermion discrete canonical-momentum values
have a functional character and read, ${\bar{q}}_j = q_j +
Q^{\Phi}_{\alpha\nu} (q_j)/L = [2\pi/ L] I^{\alpha\nu}_j +
Q^{\Phi}_{\alpha\nu} (q_j)/L$ where $j = 1, 2, ...,
N^*_{\alpha\nu}$ and $N^*_{\alpha\nu}=N_{\alpha\nu}+N_{\alpha\nu}^h$.  
Here is a $Q^{\Phi}_{\alpha,\,\nu}(q_j)/2$ is a $\alpha\nu$ pseudofermion
scattering phase shift given by,
\begin{equation}
Q^{\Phi}_{\alpha\nu} (q_j)/2 = \sum_{\alpha'\nu'}\,
\sum_{j'=1}^{N^*_{\alpha'\nu'}}\,\pi\,\Phi_{\alpha\nu,\,\alpha'\nu'}(q_j,q_{j'})\,
\Delta N_{\alpha'\nu'}(q_{j'}) \, ; \hspace{0.25cm} j = 1, 2, ...,
N^*_{\alpha\nu} \, , \label{qcan1j}
\end{equation}
where $\Delta N_{\alpha\nu}(q_{j})=\Delta {\cal{N}}_{\alpha\nu}
({\bar{q}}_j)$ is the bare-momentum distribution function
deviation $\Delta N_{\alpha\nu} (q_j) = N_{\alpha\nu} (q_j) -
N^0_{\alpha\nu} (q_j)$ corresponding to the excited energy
eigenstate. This deviation is expressed in
terms of the {\it bare-momentum} $q_j=[2\pi I^{\alpha\nu}_j]/L$,
which is carried by the $\alpha\nu$ pseudoparticles, where
$I^{\alpha\nu}_j$ are the quantum numbers provided by the
Bethe-ansatz solution \cite{I}. 

Although the $\alpha\nu$
pseudoparticles carry bare-momentum $q_j$, one can also label the
corresponding $\alpha\nu$ pseudofermions by such a bare-momentum.
When we refer to the pseudofermion bare-momentum $q_j$, we mean
that $q_j$ is the bare-momentum value that corresponds to the
pseudofermion canonical momentum ${\bar{q}}_j = q_j +
Q^{\Phi}_{\alpha\nu} (q_j)/L$. 
For the ground state the pseudofermion numbers are given by 
$N_{c0}=N$, $N_{s1}=N_{\downarrow}$, $N_{\alpha\nu}=0$ for 
$\alpha\nu\neq c0,\, s1$. We call $N^0_{c0}$ and $N^0_{s1}$ the 
ground-state $c0$ and $s1$ pseudofermion numbers, respectively. 
As mentioned in the previous section, the ground-state 
$\alpha\nu =c0,\,s1$ bare-momentum distribution functions are such
that there is pseudofermion occupancy for $\vert q\vert\leq
q^0_{F\alpha\nu}$ and unoccupancy for $q^0_{F\alpha\nu}<\vert
q\vert\leq q^0_{\alpha\nu}$, where in the thermodynamic limit the
{\it Fermi-point} values are given by,
\begin{equation}
q^0_{Fc0} = 2k_F \, ; \hspace{0.25cm} q^0_{Fs1} = k_{F\downarrow}
\, . \label{q0Fcs}
\end{equation}
Moreover, for that state the limiting bare-momentum values of both
the $\alpha\nu =c0,\,s1$ and $\alpha\nu\neq c0,\,s1$ bands read,
\begin{equation}
q^0_{c0} = \pi \, ; \hspace{0.2cm} q^0_{s1} = k_{F\uparrow} \, ;
\hspace{0.2cm} q^0_{c\nu} = [\pi -2k_F] \, , \hspace{0.1cm} \nu
>0 \, ; \hspace{0.2cm} q^0_{s\nu} =
[k_{F\uparrow}-k_{F\downarrow}] \, , \hspace{0.1cm} \nu >1 \, .
\label{qcanGS}
\end{equation}
The ground-state $\alpha\nu =c0,\,s1$ densely-packed 
bare-momentum distribution functions $N^0_{\alpha\nu} (q_j)$ 
are given in Eqs. (C.1)-(C.3) of Ref. \cite{I}. 

Under the ground-state
- excited-energy-eigenstate transitions, the $\alpha\nu$
pseudofermions and $\alpha\nu$ pseudofermion holes undergo
elementary scattering events with the $\alpha'\nu'$ pseudofermions
and $\alpha'\nu'$ pseudofermion holes created in these transitions
\cite{V-1}. This leads to the elementary two-pseudofermion phase
shifts $\pi\,\Phi_{\alpha\nu,\,\alpha'\nu'}(q_j,q_{j'})$ on the
right-hand side of the overall scattering phase shift
(\ref{qcan1j}), as further discussed in Ref. \cite{S}. 
Moreover, within the PDT the overall $\alpha\nu$ 
pseudofermion or hole phase shift,
\begin{equation}
Q_{\alpha\nu}(q_j)/2 = Q_{\alpha\nu}^0/2 + Q^{\Phi}_{\alpha\nu}
(q_j)/2 \, , \label{Qcan1j}
\end{equation}
controls the spectral weight distributions.
Here $Q_{\alpha\nu} (q_j)/L$ gives the shift in the discrete
canonical-momentum value ${\bar{q}}_j $ that arises due to the
transition from the ground state to an excited energy eigenstate
and $Q_{\alpha\nu}^0/2$ can have the values $Q_{\alpha\nu}^0/2
=0,\,\pm\pi/2$ \cite{V-1,S}. In this paper we use boundary 
conditions such that $Q_{\alpha\nu}^0/2
=0,\,-{\rm sgn}(k)\,\pi/2$, where $k$ is the excited-state
momentum relative to that of the initial ground state. Here we
assume that for the latter state $N/2$ and $N$ are odd and even
numbers, respectively. $Q_{\alpha\nu}^0/L$ gives the shift in the
discrete bare-momentum value $q_j $ that arises as a result of the
same transition. 

The $\alpha\nu$ pseudofermion creation and
annihilation operators $f^{\dag}_{{\bar{q}}_j,\,\alpha\nu}$ and
$f_{{\bar{q}}_j,\,\alpha\nu}$, respectively, have exotic
anticommutation relations \cite{V-1}. These
anticommutators involve the overall phase shifts (\ref{Qcan1j})
and play a key role in the spectral properties. There are
corresponding local operators $f^{\dag}_{x_j,\,\alpha\nu}$ and
$f_{x_j,\,\alpha\nu}$. Here $x_j$ where $j = 1, 2, ...,
N^*_{\alpha\nu}$ are the spatial coordinates of a $\alpha\nu$ 
effective lattice with $N^*_{\alpha\nu}$ sites. Each of the
$N_{\alpha\nu}$ occupied sites of such a lattice 
correspond to well-defined occupancy configurations
of $2\nu$ sites of the rotated-electron lattice. It turns out
that the operators $f^{\dag}_{x_j,\,\alpha\nu}$ and
$f_{x_j,\,\alpha\nu}$ have
simple anticommutation relations. Indeed, the exotic anticommutation relations 
of the operators and $f^{\dag}_{{\bar{q}}_j,\,\alpha\nu}$ and
$f_{{\bar{q}}_j,\,\alpha\nu}$ result from the form of the
discrete canonical momentum values ${\bar{q}}_j = q_j +
Q^{\Phi}_{\alpha\nu} (q_j)/L = [2\pi/ L] I^{\alpha\nu}_j +
Q^{\Phi}_{\alpha\nu} (q_j)/L$. 

There is a simple relation between the $f^{\dag}_{x_j,\,c0}$ and
$f_{x_j,\,c0}$ operators and those of the rotated
electrons such that the former operators have
anticommutation relations. To show that 
for branches $\alpha\nu\neq c0$ the operators
$f^{\dag}_{x_j,\,\alpha\nu}$ and
$f_{x_j,\,\alpha\nu}$ also have anticommutation relations
is a much more involved problem. The composite 
$\alpha\nu\neq c0$ pseudofermions emerge from
from corresponding $\alpha\nu\neq c0$ bond particles
through a suitable Jordan-Wigner transformation \cite{J-W}.
The point is that such $\alpha\nu\neq c0$ bond particles  
live on the corresponding $\alpha\nu$ effective lattice
and behave there as hard-core bosons. That interesting
problem will be studied in detail elsewhere. 

The charge and spin pseudofermion {\it Fermi}-point group velocities
$v_{c0}$ and $v_{s1}$ also play an important role in our study. 
The velocity $v_{\alpha\nu}$
with $\alpha\nu =c0,\,s1$ is a particular case of the 
momentum-dependent group velocity 
$v_{\alpha\nu}(q)$. Such velocities appear in all the 
spectral-weight distribution expressions of the metallic phase
and are given by,
\begin{equation}
v_{\alpha\nu}(q) = {\partial\epsilon_{\alpha\nu}(q)\over{\partial
q}} \, , \hspace{0.25cm} {\rm all}\hspace{0.15cm}{\rm branches} \,
; \hspace{0.25cm} v_{\alpha\nu}\equiv
v_{\alpha\nu}(q^0_{F\alpha\nu}) \, , \hspace{0.25cm} \alpha\nu =
c0,\,s1 \, . \label{v0}
\end{equation}
In the first expression $\epsilon_{\alpha\nu}(q)$ stands for the $\alpha\nu$
pseudofermion energy dispersion defined by Eqs. (18)-(20)
of Ref. \cite{V-1}. In this paper and its Appendix A we use often
a convention according to which the ${\bar{\alpha}\bar{\nu}}=c0,\,s1$ 
branch is that whose pseudofermion {\it Fermi}-point group velocity 
$v_{\bar{\alpha}\bar{\nu}}$ is such that,
\begin{equation}
v_{\bar{\alpha}\bar{\nu}}={\rm min}\,\{v_{c0},\,v_{s1}\}\, .
\label{convention}
\end{equation}

\section{SPECTRAL-WEIGHT DISTRIBUTIONS FOR THE METALLIC PHASE AND 
$0\leq m<n$}\label{SecIII}

\subsection{General spectral-function expressions}

The pseudofermion elementary processes that generate the PS from 
the initial ground state belong to three types:
\vspace{0.25cm}

(A) Finite-energy and finite-momentum elementary $c0$ and $s1$
pseudofermion processes away from the corresponding {\it Fermi
points} involving creation or annihilation of a finite number of
pseudofermions plus creation of $\alpha\nu\neq c0,\,s1$
pseudofermions with bare-momentum values different from the
limiting bare-momentum values $\pm q_{\alpha\nu}^0$;
\vspace{0.25cm}

(B) Zero-energy and finite-momentum processes that change the
number of $c0$ and $s1$ pseudofermions at the $\iota={\rm sgn}
(q)\,1=+1$ right and $\iota={\rm sgn} (q)\,1=-1$ left $c0$ and
$s1$ {\it Fermi points} - these processes transform the
ground-state densely packed bare-momentum occupancy configuration
into an excited-state densely packed bare-momentum occupancy
configuration. Furthermore, creation of a finite number of
independent $-1/2$ holons and independent $-1/2$ spinons,
including $-1/2$ Yang holons, $-1/2$ HL spinons, and $-1/2$ holons
and $-1/2$ spinons associated with $c\nu$ pseudofermions of
limiting bare momentum $q=\pm q_{c\nu}^0=\pm [\pi -2k_F]$ and
$s\nu$ pseudofermions of limiting bare momentum $q=\pm
q_{s\nu}^0=\pm [k_{F\uparrow}-k_{F\downarrow}]$,
respectively;\vspace{0.25cm}

(C) Low-energy and small-momentum elementary $c0$ and $s1$
pseudofermion particle-hole processes in the vicinity of the
$\iota={\rm sgn} (q)\,1=+1$ right and $\iota={\rm sgn} (q)\,1=-1$
left $c0$ and $s1$ {\it Fermi points}, relative to the
excited-state $\alpha\nu= c0,\,s1$ pseudofermion densely packed
bare-momentum occupancy configurations generated by the above
elementary processes (B).\vspace{0.25cm}

Such processes generate excitations which can be 
classified by the values of a set of numbers and number deviations. 
For instance, $N^{phNF}_{\alpha\nu}$ is the number of finite-momentum 
and finite-energy $\alpha\nu=c0,\,s1$ pseudofermion particle-hole processes of
type (A). The quantum number $\iota ={\rm sgn} (q) 1=\pm 1$ refers to the right
pseudfermion movers ($\iota =+1$) and left pseudfermion movers
($\iota =-1$) and $\Delta N^F_{\alpha\nu ,\,\iota}$ such that
$\Delta N^F_{\alpha\nu ,\,\pm 1}$ is the deviation in the number
of $\alpha\nu$ pseudofermions at the right $(+1)$ and left $(-1)$
{\it Fermi points} generated by the elementary processes (B). In turn, 
the deviation in the number of $\alpha\nu=c0,\,s1$ pseudofermions 
created or annihilated away from these points by the elementary
processes (A) is denoted by $\Delta N^{NF}_{\alpha\nu}$. 

The actual number of $\alpha\nu$ pseudofermions created or annihilated
at the right $(+1)$ and left $(-1)$ {\it Fermi points} by the
processes (B) is denoted by $\Delta N^{0,F}_{\alpha\nu ,\,\pm 1}$. 
It is such that $\Delta N^F_{\alpha\nu,\,\iota}= \Delta
N^{0,F}_{\alpha\nu,\,\iota}+\iota\,Q^ 0_{\alpha\nu}/2\pi$, where
$Q_{\alpha\nu}^0/2$ is the scattering-less phase shift on the
right-hand side of Eq. (\ref{Qcan1j}). Furthermore,
$N_{\alpha\nu,\,\iota}^{F}$ refers to the $\alpha\nu\neq c0,\, s1$
branches and is the number of $\alpha\nu$ pseudofermions of
limiting bare momentum $q=\iota\,q_{\alpha\nu}^0$ such that $\iota
=\pm 1$ created by the elementary processes (B). The number of 
$\alpha\nu$ pseudofermions created away from the limiting 
bare-momentum values by the processes (A) is called
$N_{\alpha\nu}^{NF}$.

A ground-state - excited-energy-eigenstate transition generated by 
elementary processes (A) and (B) leads to the energy and momentum 
spectrum $l\Delta E$ and $l\Delta P$ given in Eqs (28) and (29) of 
Ref. \cite{V-1}, respectively, where $l=+1$ or $l=-1$ depending on the 
specific spectral function (\ref{ABON}) under consideration. Each of such 
excited states is associated with a well defined point $(l\Delta E,l\Delta P)$ 
in the $(k,\omega)$-plane. A key property of the PDT is that the set or tower of 
excited states generated by the elementary processes (C) from each 
excited energy eigenstate generated by the processes (A) and (B)
have the same values for the two charge parameters $2\Delta_{c0}^{\pm 1}$ 
and two spin parameters $2\Delta_{s1}^{\pm 1}$ of Eq. (\ref{Delta})
as the latter state. The transitions to that tower of excited states 
generates the spectral weight in the vicinity of the 
corresponding point $(l\Delta E,l\Delta P)$. Thus, that weight
is associated with the same value of the following functional, which
plays a major role in the spectral properties,
\begin{eqnarray}
\zeta_0 & = &  2\Delta_{c0}^{+1}+ 2\Delta_{c0}^{-1}+
2\Delta_{s1}^{+1}+2\Delta_{s1}^{-1} \, ,
\nonumber \\
2\Delta_{\alpha\nu}^{\iota} & = & 
\left(\iota\,\Delta N_{\alpha\nu,\,\iota}^F+ {Q^{\Phi}_{\alpha\nu}
(\iota\,q^0_{F\alpha\nu})\over 2\pi}\right)^2 \, ,
\label{zeta0}
\end{eqnarray}
where $\alpha\nu = c0,\,s1$ and $\iota = \pm 1$.
Note that $\zeta_0$ equals the sum of the four 
parameters of Eq. (\ref{Delta}), which can be expressed in terms of the
overall scattering phase shift of Eq. (\ref{qcan1j}), as
given in the second expression of Eq. (\ref{zeta0}). Such 
parameters are functionals of the pseudofermion occupancy
configurations which describe the excited states 
generated by the elementary processes (A) and (B). 

The corresponding functional expressions are given in Eqs. (12) and (40) of 
Ref. \cite{V-1}. Thus, for each excited state generated from the initial 
ground state by the processes (A) and (B) there is a subspace spanned
by the set of excited states generated by the processes (C) from 
the former excited state. Given the linear $\alpha\nu =c0,\,s1$ 
pseudofermion energy dispersion near the {\it Fermi} points, the 
processes (C) lead to small momentum and energy values such that,
\begin{equation}
k' = \sum_{\alpha\nu=c0,\,s1}\sum_{\iota =\pm 1}\iota\,{2\pi\over
L} \,m_{\alpha\nu,\,\iota} \, ; \hspace{0.25cm} \omega' =
\sum_{\alpha\nu=c0,\,s1}\sum_{\iota =\pm 1}{2\pi\over
L}\,v_{\alpha\nu} \,m_{\alpha\nu,\,\iota} \, .
\label{DE-DP-phF-summ}
\end{equation}
Here $m_{\alpha\nu,\,\iota}$ is the number of 
elementary $\alpha\nu$ pseudofermion particle-hole processes of momentum
$\iota [2\pi/L]$. Thus, the elementary processes (C) generate a set 
of excited energy eigenstates with energy $l\Delta E$ and 
momentum $l\Delta P$ given by those of the initial excited state 
state generated from the ground state by the processes (A) and (B), 
plus the small energy $\omega'$ and momentum $k'$ provided
in Eq. (\ref{DE-DP-phF-summ}).

Within the PDT the general $\cal{N}$-electron spectral functions
(\ref{ABON}) factorize in terms of $\alpha\nu$ pseudofermion spectral
functions. The probability amplitude $A^{(0,0)}_{\alpha\nu}$ appears 
in the expressions of the latter functions for $\alpha\nu =c0,\,s1$.
It is associated with the canonical-momentum densely packed 
configurations generated by the elementary processes (A) and (B) 
after the matrix elements between the ground state and the excited 
states generated by these processes are computed. Such a probability 
amplitude corresponds to the $\alpha\nu =c0,\,s1$ pseudofermion
spectral-function lowest-peak weight given in Eq. (48) of Ref.
\cite{V-1}. It has the following approximate behaviour,
\begin{eqnarray}
A^{(0,0)}_{\alpha\nu} & \approx & \prod_{\iota =\pm 1}
A^{(0,0)}_{\alpha\nu,\,\iota}\Bigl[1+{\cal{O}}\Bigl({1\over
N_a}\Bigr)\Bigr] \, , 
\nonumber \\
A^{(0,0)}_{\alpha\nu,\,\iota}
& = & {f_{\alpha\nu,\,\iota} \over \Bigl(N_a\,
S^0_{\alpha\nu}\Bigr)^{-1/2+ 2\Delta_{\alpha\nu}^{\iota}}} \, ;
\hspace{0.25cm} \alpha\nu = c0,\,s1 \, ; \hspace{0.15cm} \iota =
\pm 1 \, . \label{A002}
\end{eqnarray}
Here $2\Delta_{\alpha\nu}^{\iota}$ is the functional given in Eq.
(\ref{Delta}), and $f_{\alpha\nu,\,\iota}$ reads,
\begin{equation}
f_{\alpha\nu,\,\iota} = \sqrt{f\Bigl(Q_{\alpha\nu}(\iota
q^0_{F\alpha\nu})+{\rm sgn} (k)\pi\Bigr)} \, ; \hspace{0.2cm}
f_{\alpha\nu} = \prod_{\iota =\pm 1}f_{\alpha\nu,\,\iota} \, ;
\hspace{0.1cm} \alpha\nu = c0,\,s1 \, . \label{fan}
\end{equation}
In this expression, $k$ stands for the excited-state momentum relative 
to the initial ground state, $f (Q)=f(-Q)$ is the function defined in 
Ref. \cite{Karlo}, which occurs on the right-hand 
side of Eq. (24) of that reference, and $f_{\alpha\nu}$ appears in spectral-function
expressions introduced below. Moreover, $S^0_{\alpha\nu}$ is a
$n$, $m$, and $U/t$ dependent constant such that
$S^0_{c0}\,S^0_{s1}\rightarrow 1$ both for $U/t\rightarrow 0$ and
for $U/t\rightarrow \infty$ and $m\rightarrow 0$. (From 
Ref. \cite{Karlo} we learn that $S^0_{c0}\rightarrow\sin (\pi n)$ for
$U/t\rightarrow \infty$ and $m\rightarrow 0$, and thus
$S^0_{s1}\rightarrow 1/\sin (\pi n)$ in such a limit.) It is
useful to introduce the following quantity,
\begin{equation}
D_0 = \prod_{\alpha\nu =
c0,\,s1}{S^0_{\alpha\nu}\,f_{\alpha\nu}\over
(S^0_{\alpha\nu})^{[2\Delta^{+1}_{\alpha\nu}+2\Delta^{-1}_{\alpha\nu}]}}  \, .
\label{D0}
\end{equation}

Another important peace of the $\alpha\nu =c0,\,s1$ pseudofermion
spectral functions is the relative weight 
$a_{\alpha\nu,\,\iota}(m_{\alpha\nu,\,\iota})$ generated by the elementary 
processes (C) in the vicinity of the right ($\iota =+1$) and left ($\iota =-1$) 
{\it Fermi} points. When $2\Delta_{\alpha\nu}^{\iota}>0$ the
weight $a_{\alpha\nu,\,\iota}(m_{\alpha\nu,\,\iota})$ and its
asymptotic expression are given in Eqs. (52) and (55) of Ref. \cite{V-1},
respectively. The point is that when the excited states generated by 
the processes (A) and (B) are such that $2\Delta_{\alpha\nu}^{\iota}=0$
the relative weight $a_{\alpha\nu,\,\iota}(m_{\alpha\nu,\,\iota})$ reads instead,
\begin{equation}
a_{\alpha\nu,\,\iota} (m_{\alpha\nu,\,\iota}) = \delta_{m_{\alpha\nu,\,\iota},0}
\, ; \hspace{0.50cm} 2\Delta_{\alpha\nu}^{\iota}= 0 \, ;
\hspace{0.25cm} \alpha\nu = c0,\,s1 \, ; \hspace{0.25cm} \iota =
\pm 1 \, . \label{aD=0}
\end{equation}
Since for densities in the range $0<n<1$ and $0<m<n$ 
the two charge parameters $2\Delta_{c0}^{\pm 1}$ and two spin 
parameters $2\Delta_{s1}^{\pm 1}$ are finite for all
excited states, this case was not considered in the studies of 
Ref. \cite{V-1}. 

As further discussed below, the general $\cal{N}$-electron spectral function
given in Eq.  (\ref{ABON}) can be expressed in terms of the charge $c0$ and 
spin $s1$ spectral functions provided in Eq. (56) of Ref. \cite{V-1}. Here we 
express the latter functions as follows,
\begin{eqnarray}
B^{l,i}_{Q_{\alpha\nu}} (k',\omega') & = & {N_a\over
2\pi}\int_{-\infty}^{+\infty} dk''\int_{-\infty}^{+\infty}
d\omega''\,B^{l,\iota,i}_{Q_{\alpha\nu}} \Bigl(k'-k'',\omega'
-\omega''\Bigr)\nonumber \\
& \times & B^{l,-\iota,i}_{Q_{\alpha\nu}} \Bigl(k'',\omega''\Bigr) \, ; 
\hspace{0.25cm} \alpha\nu = c0, s1\, ;
\hspace{0.25cm} \iota = \pm 1 \, . \label{BQBQ}
\end{eqnarray}
The studies of Ref. \cite{V-1} refer to the case where
the two charge parameters $2\Delta_{c0}^{\pm 1}$ and two spin 
parameters $2\Delta_{s1}^{\pm 1}$ are finite. In that
case the four relative weights $a_{\alpha\nu,\,\iota}(m_{\alpha\nu,\,\iota})$ 
such that $\alpha\nu = c0, s1$ and $\iota = \pm 1$ and their
asymptotic expressions are of the form provided in Eqs. (52) and 
(55) of Ref. \cite{V-1}, respectively, and thus the functions 
$B^{l,\iota,i}_{Q_{\alpha\nu}} (k',\omega')$ on the right-hand 
side of Eq. (\ref{BQBQ}) are given by \cite{V-1},
\begin{eqnarray}
& & B^{l,\iota,i}_{Q_{\alpha\nu}} (k',\omega') =
{A^{(0,0)}_{\alpha\nu,\,\iota}\over v_{\alpha\nu}}\,
\,a_{\alpha\nu,\,\iota}
\Bigl({l\,\omega'\over 2\pi
v_{\alpha\nu}/N_a}\Bigr)\,\delta (k'-{\iota\,\omega'\over
v_{\alpha\nu}}) \nonumber \\
& \approx & 
{f_{\alpha\nu,\,\iota}\over
v_{\alpha\nu}\,\Gamma
(2\Delta_{\alpha\nu}^{\iota})}\,
{\Theta (l\omega')\over \sqrt{N_a\,S^0_{\alpha\nu}}}\,
\Bigr({l\omega'\over 2\pi\,
v_{\alpha\nu}\,S^0_{\alpha\nu}}\Bigl)^{2\Delta_{\alpha\nu}^{\iota}-1}\,
\delta \Bigl(k'-{\iota\,\omega'\over v_{\alpha\nu}}\Bigr) \, ,
\label{Bkom-sum-G}
\end{eqnarray}
where $\alpha\nu = c0,\, s1$ and $\iota =\pm 1$.
The second expression corresponds to the asymptotic behaviour
valid for small finite values of $l\omega'$.

However, when $2\Delta_{\alpha\nu}^{\iota}= 0$ the corresponding
relative weight $a_{\alpha\nu,\,\iota}(m_{\alpha\nu,\,\iota})$
is of the form provided in Eq. (\ref{aD=0}) and thus the spectral 
function $B^{l,\iota,i}_{Q_{\alpha\nu}} (k',\omega')$ instead
of being given by Eq. (\ref{Bkom-sum-G}) reads,
\begin{eqnarray}
B^{l,\iota,i}_{Q_{\alpha\nu}} (k',\omega') & = &
{2\pi\over N_a}A^{(0,0)}_{\alpha\nu,\,\iota}\,\delta (k')\,\delta (\omega') \nonumber \\
& \approx & 
2\pi\,f_{\alpha\nu,\,\iota}\,\sqrt{S^0_{\alpha\nu}\over N_a}
\,\delta (k')\,\delta (\omega')
\, ; \hspace{0.25cm} \alpha\nu = c0,\, s1 \, ;
\hspace{0.25cm} \iota = \pm 1 \, . \label{Bkom-sum-G-D0}
\end{eqnarray}
\begin{figure}
\includegraphics[scale=.8]{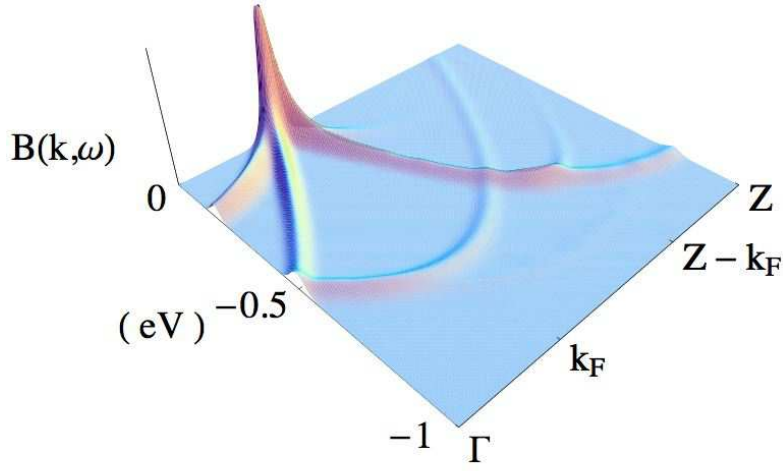} 
\caption{\label{fig1}
Full theoretical line shape of the one-electron 
removal spectral-weight distribution found in Ref. \cite{TTF}
to fit the corresponding spectral features of TTF-TCNQ.
The spectrum shown here is rotated relative to that
shown in Fig. 2 of that reference. It includes both 
the TTF related spectral features for 
$n=1.41;t = 0.35\hspace{0.1cm}{\rm eV};U/t=5.61$ and 
those of TCNQ for $n=0.59;t=0.40\hspace{0.1cm}{\rm eV};U/t=4.90$, 
respectively.}
\end{figure}

In order to reach the correct spectral-function expressions for
all electronic densities of the metallic phase and for 
$m\rightarrow 0$, several cases must be considered. 
The first corresponds to excited states such that the two charge 
parameters $2\Delta_{c0}^{\pm 1}$ and two spin 
parameters $2\Delta_{s1}^{\pm 1}$ are finite. The two
functions on the right-hand side of Eq. (\ref{BQBQ}) are of the
form given in Eq. (\ref{Bkom-sum-G}) and thus we find,
\begin{eqnarray}
& & B^{l,i}_{Q_{\alpha\nu}} (k',\omega') = {L\over 4\pi
v_{\alpha\nu}}\,A^{(0,0)}_{\alpha\nu}\,\prod_{\iota =\pm
1}\,a_{\alpha\nu,\,\iota}\Bigl({l[\omega'
+\iota\,v_{\alpha\nu}\,k']\over 4\pi v_{\alpha\nu}/N_a}\Bigr) \nonumber \\
& \approx &   
{f_{\alpha\nu}\over
4\pi\,v_{\alpha\nu}\,S^0_{\alpha\nu}}\,\prod_{\iota =\pm
1}\,{\Theta (l[\omega' +\iota\,v_{\alpha\nu}\,k'])\over 
\Gamma (2\Delta_{\alpha\nu}^{\iota})}\,\Bigl({l[\omega'
+\iota\,v_{\alpha\nu}\,k']\over 4\pi
\,v_{\alpha\nu}\,S^0_{\alpha\nu}}\Bigr)^{2\Delta_{\alpha\nu}^{\iota}-1} \, ,
\label{B-J-i-sum-GG}
\end{eqnarray}
where $\alpha\nu = c0,\, s1$.
The second expression corresponds to the asymptotic behaviour
valid for small finite values of $l[\omega' +\iota\,v_{\alpha\nu}\,k']$.
This function is that already given in Eq. (56) of Ref. \cite{V-1}.
(See the note \cite{note} concerning a discrepancy of a factor
$2$ between the function given here and that provided
in Eq. (56) of Ref. \cite{V-1} and how a misprint in the
latter equation becomes an error, by propagating 
to other expressions of Refs. \cite{V-1,LE}.)

The second case occurs when the excited states are such that
for the functions on the right-hand side of Eq. (\ref{BQBQ})
one has that $2\Delta_{\alpha\nu}^{\iota}>0$ and
$2\Delta_{\alpha\nu}^{-\iota}=0$. Then one must use in that
equation the expression (\ref{Bkom-sum-G}) for the function
$B^{l,\iota,i}_{Q_{\alpha\nu}} (k',\omega')$ and that given
in Eq. (\ref{Bkom-sum-G-D0}) for the function
$B^{l,-\iota,i}_{Q_{\alpha\nu}} (k',\omega')$. This leads
to the following expression for the function on the 
left-hand side of Eq. (\ref{BQBQ}),
\begin{eqnarray}
& & B^{l,i}_{Q_{\alpha\nu}} (k',\omega') =
{A^{(0,0)}_{\alpha\nu}\over v_{\alpha\nu}}\,
\,a_{\alpha\nu,\,\iota}
\Bigl({l\,\omega'\over 2\pi
v_{\alpha\nu}/N_a}\Bigr)\,\delta \Bigl(k'-{\iota\,\omega'\over v_{\alpha\nu}}\Bigr) 
\nonumber \\
& \approx & {f_{\alpha\nu}\over
v_{\alpha\nu}\,\Gamma
(2\Delta_{\alpha\nu}^{\iota})}\,\Theta (l\omega')
\Bigr({l\omega'\over 2\pi\,
v_{\alpha\nu}\,S^0_{\alpha\nu}}\Bigl)^{2\Delta_{\alpha\nu}^{\iota}-1}\,
\delta \Bigl(k'-{\iota\,\omega'\over v_{\alpha\nu}}\Bigr) \, , 
\label{B-J-i-sum-GG-D0}
\end{eqnarray}
where $\alpha\nu = c0,\, s1$ and $\iota = \pm 1$.
\begin{figure}
\includegraphics[scale=0.8]{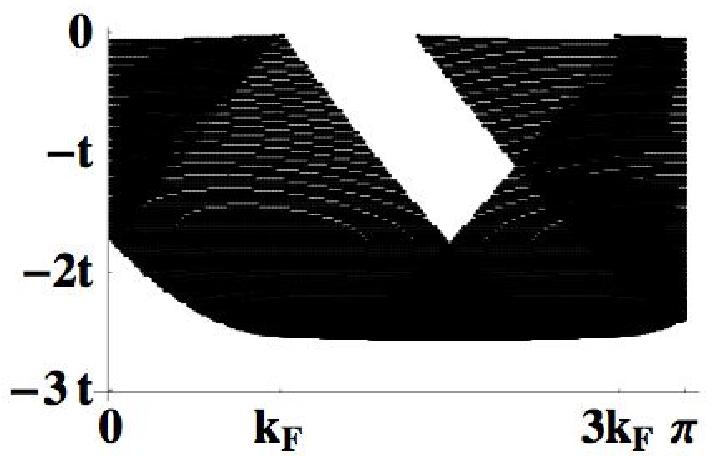} 
\includegraphics[scale=0.8]{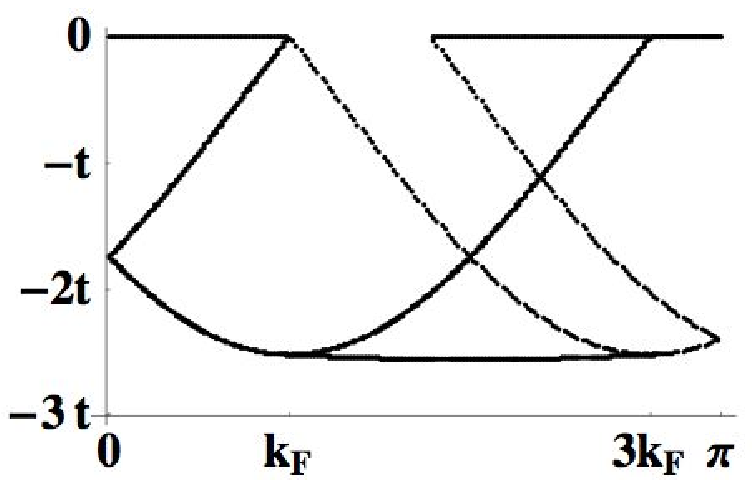}
\caption{\label{fig2}
The region of the $(k,\omega)$ plane with a finite one-electron
removal spectral weight obtained by running over all the momentum 
values of the $c0$ and $s1$ momentum distribution
deviations of Eqs. (\ref{DNq-ER1s1H}) and (\ref{DNqm0-ER1s1H})
for $U/t=100$, $n=0.59$, and $m\rightarrow 0$.
The corresponding branch and border lines are also
plotted.}
\end{figure}

Finally, the third case refers to excited states such that for 
the functions on the right-hand side of Eq. (\ref{BQBQ})
one has that $2\Delta_{\alpha\nu}^{\iota}=2\Delta_{\alpha\nu}^{-\iota}=0$.
In this case one must use in that equation the expression 
(\ref{Bkom-sum-G-D0}) for both the functions
$B^{l,\iota,i}_{Q_{\alpha\nu}} (k',\omega')$ and 
$B^{l,-\iota,i}_{Q_{\alpha\nu}} (k',\omega')$. One then
reaches the following expression for the function on the 
left-hand side of Eq. (\ref{BQBQ}),
\begin{eqnarray}
B^{l,i}_{Q_{\alpha\nu}} (k',\omega') & = &
{2\pi\over N_a}A^{(0,0)}_{\alpha\nu}\,\delta (k')\,\delta (\omega') 
\nonumber \\
& \approx & 2\pi\,f_{\alpha\nu}\,S^0_{\alpha\nu}\,\delta (k')\,\delta (\omega')
\, ; \hspace{0.25cm} \alpha\nu = c0,\, s1 \, . \label{B-J-i-sum-GG-DD0}
\end{eqnarray}

The general PDT expression for the $\cal{N}$-electron spectral function
given in Eq.  (\ref{ABON}) reads \cite{V-1},
\begin{equation}
B_{{\cal{N}}}^{l} (k,\,\omega) =
\sum_{i=0}^{\infty}c^l_i\sum_{\{\Delta
N_{\alpha\nu}\},\,\{L_{\alpha,\,-1/2}\}}\Bigl[\sum_{\{N^{phNF}_{\alpha\nu}\},\,\{\Delta
N^F_{\alpha\nu ,\,\iota}\},\,\{N^F_{\alpha\nu
,\,\iota}\}}\,B^{l,i} (k,\,\omega)\Bigr] \, , 
\label{ABONjl-J-CPHS}
\end{equation}
where $c^l_0= 1$, $l=\pm 1$, and 
for densities $0<n<1$ the summation over the numbers 
$N^{phNF}_{\alpha\nu}$ is limited to finite values. On
the right-hand side of Eq. (\ref{ABONjl-J-CPHS}), $c^l_i$ is the
constant of the operator expressions given in Eqs. (32)-(34) of
Ref. \cite{V} such that $c^l_i\rightarrow 0$ as
$U/t\rightarrow\infty$ for $i>0$ and the function $B^{l,i}
(k,\,\omega)$ is defined in Eq. (44) of Ref. \cite{V-1}. The latter
function is fully defined by the related function 
${\breve{B}}^{l,i} (\Delta\omega,v)$. However, the second expression 
for ${\breve{B}}^{l,i} (\Delta\omega,v)$ given in Eq. (45) of Ref. \cite{V-1} 
is only valid for $v_{c0}>v_{s1}$. For finite values of $U/t$ this excludes 
electronic densities in the vicinity of $1$. 
\begin{figure}
\includegraphics[scale=0.8]{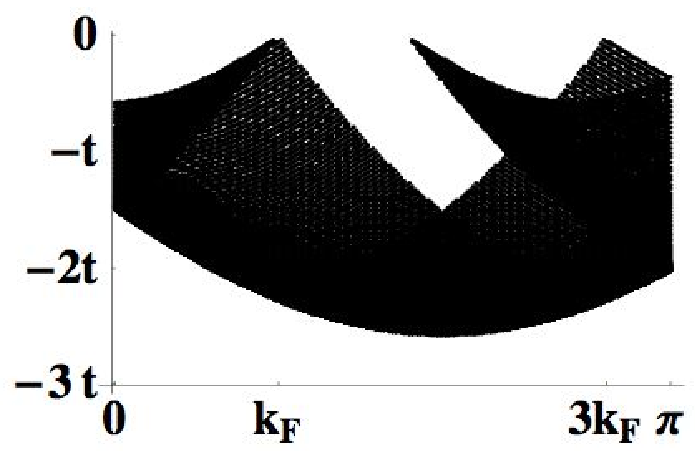} 
\includegraphics[scale=0.8]{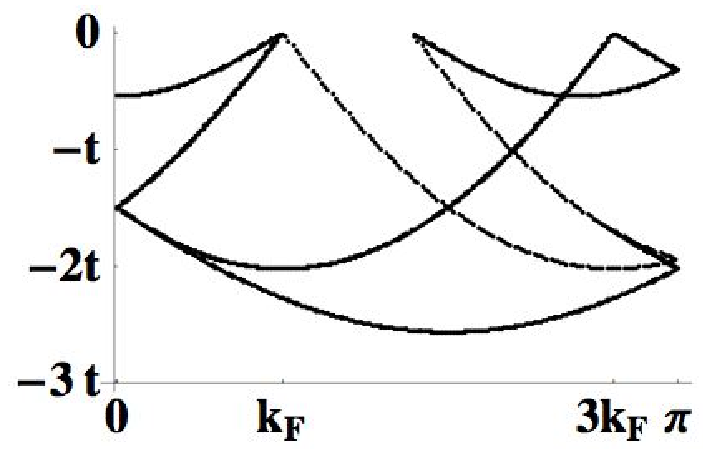}
\caption{\label{fig3}
The same region of the $(k,\omega)$ plane and branch and border
lines as in Fig. 2 for $U/t=4.9$, $n=0.59$, and $m\rightarrow 0$.}
\end{figure}

Generalization of ${\breve{B}}^{l,i} (\Delta\omega,v)$ for the whole 
range of electronic densities of the metallic phase, including densities
such that the spin and charge velocities obey the inequality 
$v_{s1}>v_{c0}$, leads to the following expression,
\begin{eqnarray}
{\breve{B}}^{l,i} (\Delta\omega,v) & = & {{\rm sgn} (v)\over
2\pi}\int_{0}^{\Delta\omega}d\omega'\int_{-{\rm sgn}
(v)\Delta\omega/v_{\alpha\nu}}^{+{\rm sgn} (v)\Delta\omega/v_{\alpha\nu}}dk'
\nonumber \\
& \times & B^{l,i}_{Q_{{\bar{\alpha}\bar{\nu}}}} \Bigl(\Delta\omega/v -
k',\Delta\omega-\omega'\Bigr)\,B^{l,i}_{Q_{\alpha\nu}}
\Bigl(k',\omega'\Bigr) \, . 
\label{B-l-i-breve}
\end{eqnarray}
In this equation, in those provided in Appendix A, and in the whole our analysis 
until subsection 3.2 we use 
a notation for the charge branch $c0$ and spin branch $s1$ such that
$\alpha\nu =c0,\,s1$ when the branch index ${\bar{\alpha}\bar{\nu}}$
defined by Eq. (\ref{convention}) reads ${\bar{c}\bar{0}}=s1,\,c0$,
respectively. Moreover, in equation (\ref{B-l-i-breve}) $l=+1$ or $l=-1$ according 
to the $\cal{N}$-electron spectral function (\ref{ABON}) under consideration, 
$\Delta\omega=(\omega-l\Delta E)$, 
$\Delta k=(k -l\Delta P)$, and $l\Delta E$ and $l\Delta P$ correspond to the 
general energy and momentum spectra given in Eqs. (28) and (29) of Ref. 
\cite{V-1}, respectively, which is generated by the elementary processes
(A) and (B), $i=0,1,2,...$, and for $i>0$ the index $i=1,2,...$ is a
positive integer number which increases for increasing values of the number of extra
pairs of creation and annihilation rotated-electron operators in the expressions of the
pseudofermion operators associated with the function 
${\breve{B}}^{l,i} (\Delta\omega,v)$ relative to that of the pseudofermion operators 
of the $i=0$ function ${\breve{B}}^{l,0} (\Delta\omega,v)$ \cite{V-1}. 
Expression (\ref{B-l-i-breve}) is valid for the whole $(k,\,\omega)$-plane, except 
for $k$ and $\omega$ values such that 
$\omega\approx \iota\,v_{\alpha\nu}(k-lk_0)+l\omega_0$, where
$\alpha\nu = c0,\,s1$, $\iota =\pm 1$, $\omega_0$ and $k_0$ are provided in
Eqs. (32) and (34) of Ref. \cite{V-1}.

The velocity $v$ appearing in the argument of the function 
(\ref{B-l-i-breve}) plays an important role in our study and is given by,
\begin{equation}
v = {\Delta\omega\over \Delta k} = {(\omega-l\Delta E)\over
(k-l\Delta P)} \, ; \hspace{1cm} {\rm sgn} (v)\,1 = {\rm sgn}
(\Delta k)\,l \, ; \hspace{1cm} \vert v\vert
>v_{\bar{\alpha}\bar{\nu}}\, . \label{V-ok}
\end{equation}
The inequality $\vert v\vert >v_{\bar{\alpha}\bar{\nu}}$ follows from the structure of the spectrum 
of Eq. (31) of Ref. \cite{V-1}. 
\begin{figure}
\includegraphics[scale=0.8]{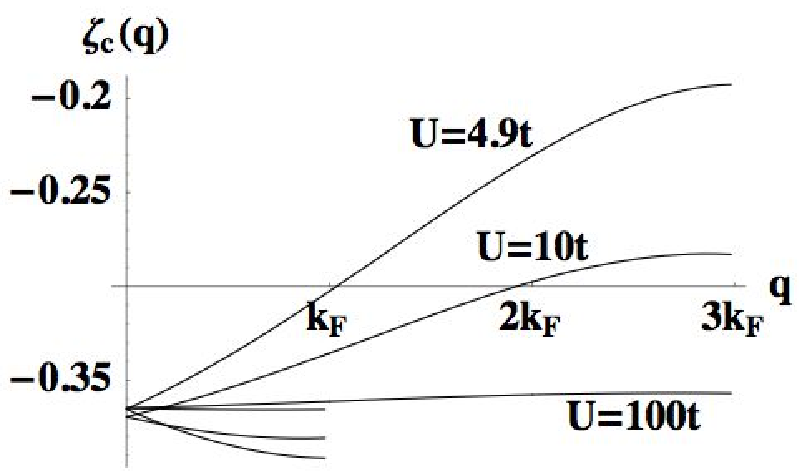} 
\includegraphics[scale=0.8]{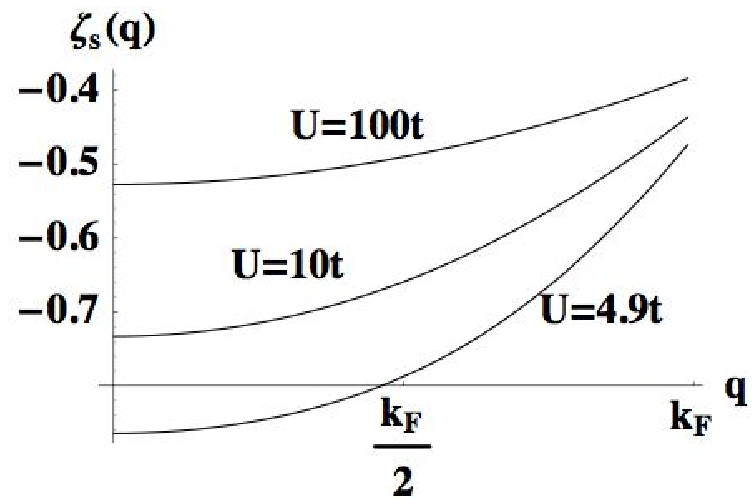}
\caption{\label{fig4}
The value of the $c0$ and $s1$ branch lines 
exponents of the one-electron removal weight
distribution as a function of momentum for 
$n=0.59$, $m\rightarrow 0$, and several magnitudes
of $U/t$. The exponents of index $c$ and $s$ refer to
the $c0$ and $s1$ branch lines, respectively.}
\end{figure}

When at least one 
of the two parameters $2\Delta_{\alpha\nu}^{\pm 1}$ 
and two parameters $2\Delta_{\bar{\alpha}\bar{\nu}}^{\pm 1}$ is finite,
use of the general expression (\ref{B-l-i-breve}) for the function
${\breve{B}}^{l,i} (\Delta\omega,v)$ for small finite values of
$l\Delta\omega =l(\omega -l\Delta E)$ with $B^{l,i}_{Q_{\alpha\nu}}(k',\omega')$ and
$B^{l,i}_{Q_{\bar{\alpha}\bar{\nu}}}(k',\omega')$ equaling the suitable
expressions provided in Eqs. (\ref{B-J-i-sum-GG}),  
(\ref{B-J-i-sum-GG-D0}), and (\ref{B-J-i-sum-GG-DD0})
leads to the following asymptotic behaviour for that
convolution function,
\begin{equation}
{\breve{B}}^{l,i} (\Delta\omega,v) \approx {F_0 (1/v)\over
4\pi\sqrt{v_{c0}\,v_{s1}}}\,\Theta
\Bigl(l\Delta\omega\Bigr)\,\Bigl({l\Delta\omega\over
4\pi\sqrt{v_{c0}\,v_{s1}}} \Bigr)^{-2+\zeta_0} \, .
\label{B-breve-asym}
\end{equation}
Here $i=0,1,2,...$, $l= \pm 1$, and $\zeta_0$ is the functional provided in Eq. (\ref{zeta0}) and
the pre-factor function $F_0 (z)$ is given in Appendix A. In turn,
if all four above parameters vanish we find by use of expression 
(\ref{B-l-i-breve}) with $B^{l,i}_{Q_{\alpha\nu}}(k',\omega')$ and
$B^{l,i}_{Q_{\bar{\alpha}\bar{\nu}}}(k',\omega')$ of the
form given in Eq. (\ref{B-J-i-sum-GG-DD0}) that,
\begin{equation}
{\breve{B}}^{l,i} (\Delta\omega,v) = {\bar{B}}^{l,i} (\Delta\omega,\Delta k)
= 2\pi D_0\,\delta (\Delta\omega)\delta (\Delta k) \, ,
\label{all-zero}
\end{equation}
where $D_0$ is defined in Eq. (\ref{D0}), in agreement with the first 
expression of Eq. (59) of Ref. \cite{V-1} for $\zeta_0 =0$.

Depending on the values of the two parameters $2\Delta_{\alpha\nu}^{\pm 1}$ 
and two parameters $2\Delta_{\bar{\alpha}\bar{\nu}}^{\pm 1}$, we 
consider in Appendix A seven cases where by use of Eqs. (\ref{B-J-i-sum-GG}),  
(\ref{B-J-i-sum-GG-D0}), (\ref{B-J-i-sum-GG-DD0}), (\ref{B-l-i-breve}),
and (\ref{B-breve-asym}) a set of alternative expressions for the pre-factor 
function $F_0 (z)$ on the right-hand side of the latter equation can
be derived and are given in that Appendix .
Expressions (\ref{C-v})-(\ref{C-v-3D0}) of the same Appendix provide a generalization of
the pre-factor function $F_0 (z)$ given in Eq. (62) of Ref. \cite{V-1}.
The latter function corresponds to the general function (\ref{C-v})
of Appendix A for $\alpha\nu =c0$ \cite{note}. The recent studies of Ref. \cite{TTF} on the finite-energy
spectral features of the organic compound TTF-TCNQ use the functions 
(\ref{C-v})-(\ref{C-v-3D0}) of that Appendix derived in this paper.
\begin{figure}
\includegraphics[scale=.8]{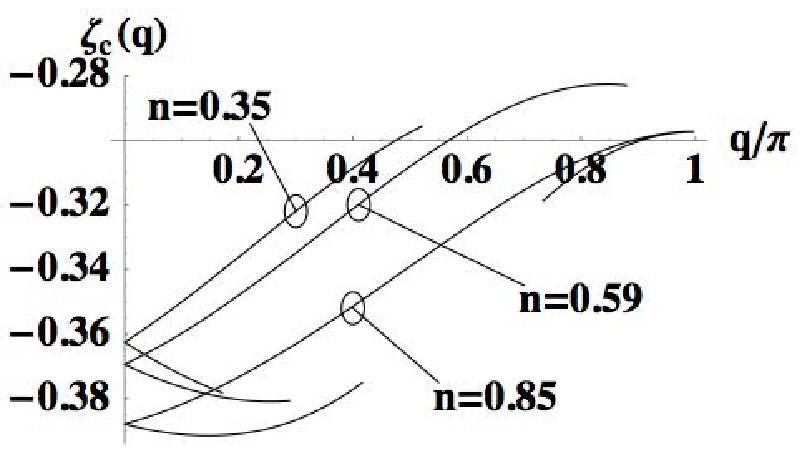} 
\includegraphics[scale=0.8]{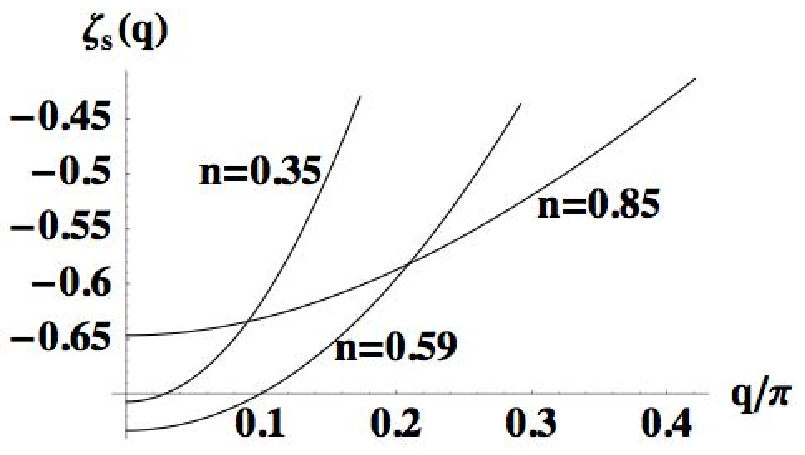} 
\caption{\label{fig5}
The same exponents as in Fig. 4 as a function of momentum for 
$U/t=10$, $m\rightarrow 0$, and several values of $n$.}
\end{figure}

Importantly, general spectral-function expressions for all electronic 
densities of the metallic phase are obtained by replacing the spin velocity $v_{s1}$ by
$v_{{\bar{\alpha}\bar{\nu}}}$ in the limits of the variable $z$ integrations
of expressions (66) and (B.14) of Ref. \cite{V-1} and both in the limits of the 
variable $z$ integrations and arguments of the theta-functions of 
expressions (71), (B.17), and (B.18) of the same reference. 
After such replacements, Eqs. (66), (68), (70), and (71) of 
Ref. \cite{V-1} with the pre-factor function $F_0 (z)$ given in 
Eqs. (\ref{C-v})-(\ref{C-v-3D0}) of Appendix A provide general spectral-function
expressions for all electronic densities of the metallic phase and spin densities
in the range $0\leq m<n$. However, note that while the branch 
index $\bar{\alpha}\bar{\nu}$ is that defined by Eq. (\ref{convention}),
the branch indices $\alpha\nu$ and $\alpha'\nu'$ of Eqs. (66) and 
(B.14) of Ref. \cite{V-1} correspond to the two created quantum 
objects and in contrast to the branch index $\alpha\nu$ of
Eqs. (\ref{B-l-i-breve})-(\ref{all-zero}) and
(\ref{C-v})-(\ref{C-v-3D0}) of Appendix A
there are no restrictions imposing that 
such branch indices are different or equal to the branch index 
$\bar{\alpha}\bar{\nu}$, the same applying to the branch index 
$\alpha\nu$ of Eqs. (70), (71), (B.17), and (B.18) of that reference. 

It is straightforward to show that the double Fourier transform 
of the function (\ref{Bkom-sum-G-D0}) equals that of the function
(\ref{Bkom-sum-G}) as $2\Delta_{\alpha\nu}^{\iota}\rightarrow 0$.
It follows that the asymptotic expression of the correlation function
given in Eq. (32) of Ref. \cite{LE} is of the general form provided
in Eq. (52) of that reference independently on whether all four 
parameters $2\Delta_{\alpha\nu}^{\iota}$ where $\alpha\nu =c0,s1$
and $\iota =\pm 1$ are finite or some of
these parameters vanish. Thus, the different expressions 
(\ref{C-v})-(\ref{C-v-3D0}) of Appendix A for the pre-factor $F_0(z)$ of the
convolution function (\ref{B-breve-asym}) are not associated
with different pre-factors for the corresponding correlation-function expression
(52) of Ref. \cite{LE}. The pre-factor of the latter function is
always of the form given in Eq. (50) of that reference \cite{note}.
\begin{figure}
\includegraphics[scale=0.8]{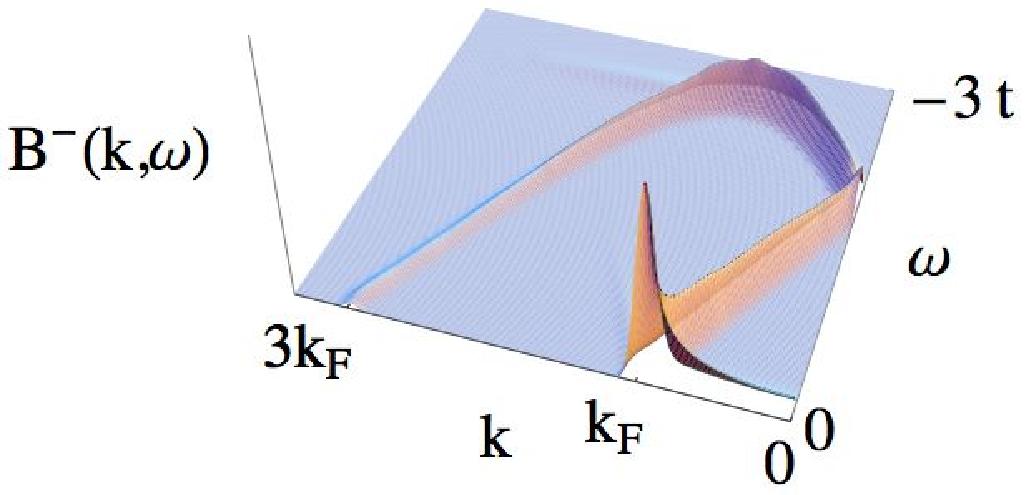} 
\includegraphics[scale=0.8]{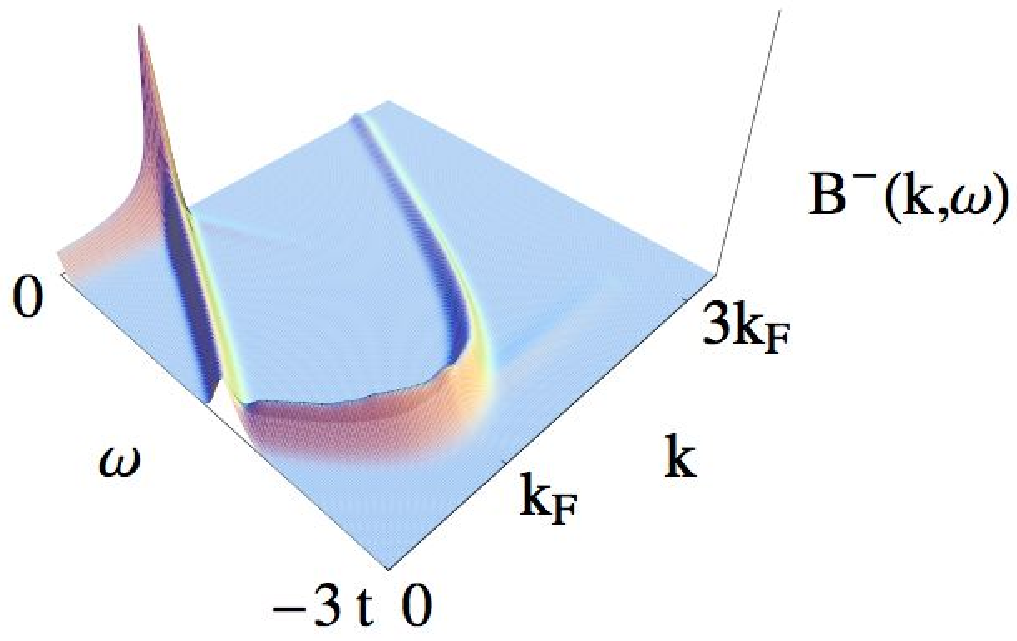}
\caption{\label{fig6}
Full PDT one-electron removal spectral-weight distribution 
associated with the $c0$ and $s1$ momentum deviations of 
Eqs. (\ref{DNq-ER1s1H}) and (\ref{DNqm0-ER1s1H}) 
and corresponding functionals given in 
Eqs. (\ref{Qc0-ER1s1H}) and (\ref{Qs1-ER1s1H})
for $U/t=100$, $n=0.59$, and $m\rightarrow 0$.}
\end{figure}

\subsection{Explicit spectral-function expressions in the vicinity of the border lines}

The singular features of the $\cal{N}$-electron spectral functions
(\ref{ABON}) are of power law type. The power-law branch-line
singular features of general form given in Eqs. (68) and (71) of Ref. \cite{V-1} are
controlled by non-classic interaction, density, and momentum dependent exponents. The 
spectral feature of general form given in Eqs. (66) of that 
reference can also include power-law singular features
called border lines. However, the studies of Ref. \cite{V-1}
did not provide an explicit general expression for the
$k$ and $\omega$ dependence in the vicinity the border
lines. Here we provide such an expression and find that
for the border lines the power-law exponent has an universal
value given by $-1/2$. 
 
The shape $\omega =\omega_{BL} (k)$ of a border line is defined 
by the following parametric equations \cite{V-1},
\begin{eqnarray}
\omega_{BL} (k) & = & l[\,\omega_0 +c_1'\,\epsilon_{\alpha'\nu'}
(q')+c_1''\,\epsilon_{\alpha''\nu''} (q'')]\,\delta_{v_{\alpha'\nu'}
(q'),\,v_{\alpha''\nu''} (q'')} \, , 
\nonumber \\
k & = & lk_{\alpha'\nu',\,\alpha''\nu''}(q',\,q'')\,\delta_{v_{\alpha'\nu'}
(q'),\,v_{\alpha''\nu''} (q'')} \, . \label{b-lines}
\end{eqnarray}
Here $q'$ and $q''$ are the bare-momentum values of the
$\alpha'\nu'$ and $\alpha''\nu''$, respectively, pseudofermion
or pseudofermion-hole scattering centers created by
the processes (A), $c_1',c_1''=+1$ for pseudofermion
creation and $c_1',c_1''=-1$ for pseudofermion-hole
creation, $\omega_0$ is the finite-energy parameter 
given in Eq. (32) of Ref. \cite{V-1}, and $k_{\alpha'\nu',\,\alpha''\nu''}(q',\,q'')$
is the momentum spectrum provided in Eqs. (64) and
(65) of that reference. As for Eqs. (66) and 
(B.14) of Ref. \cite{V-1} and in contrast to the branch index 
$\alpha\nu$ of Eqs. (\ref{B-l-i-breve})-(\ref{C-v-3D0}), there are no 
restrictions imposing that the branch indices $\alpha'\nu'$ and $\alpha''\nu''$
of the two created scattering centers are different or equal to the 
branch index $\bar{\alpha}\bar{\nu}$ defined by Eq. (\ref{convention}).
\begin{figure}
\includegraphics[scale=.8]{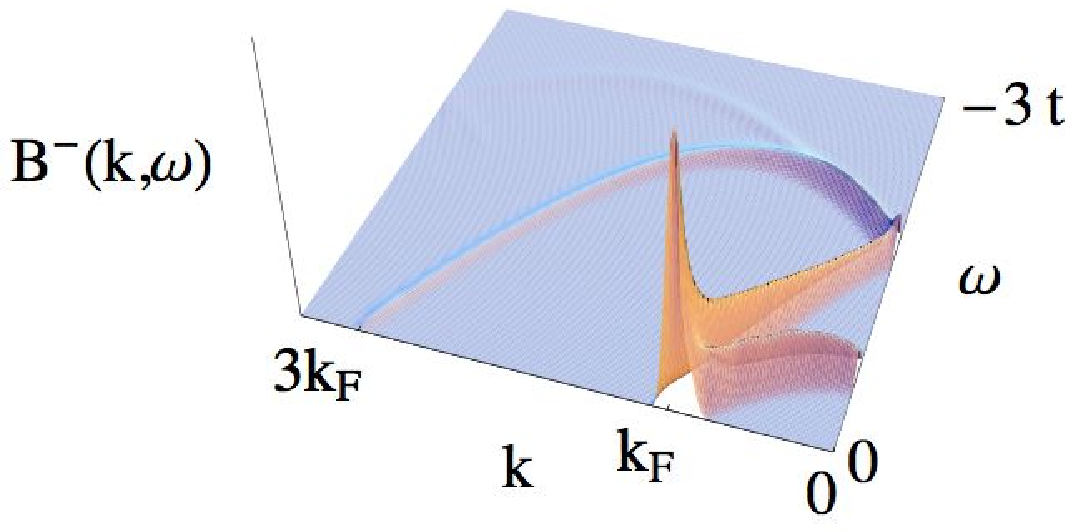} 
\includegraphics[scale=0.8]{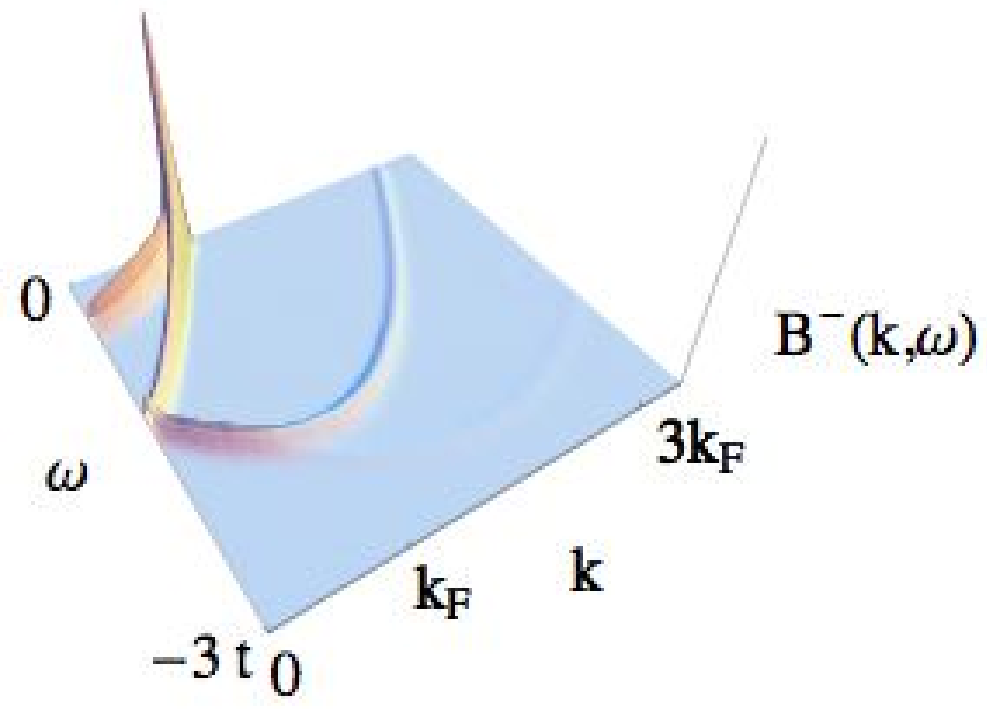} 
\caption{\label{fig7}
The same PDT one-electron removal spectral-weight distribution 
as in Fig. 6 for the values $U/t=4.9$, $n=0.59$, and 
$m\rightarrow 0$ suitable to the TCNQ related spectral
features.}
\end{figure}

By use of methods similar to those used in Ref. \cite{V-1} for 
other weight distributions, we find that the spectral function has the
following singular behaviour in the vicinity and just below
($l=+1$) or above ($l=-1$) the border line,
\begin{eqnarray}
B_{{\cal{N}}}^{l} (k,\,\omega) & \approx & {2\,\Theta \Bigl(\Omega
-l[\omega_{BL} (k)-\omega]\Bigr)\over \pi\,C_c\,C_s\,\zeta_0
(q',\,q'')\,\sqrt{\vert a_{\alpha'\nu'} (q')\vert +\vert
a_{\alpha''\nu''} (q'')\vert}}
\nonumber \\
& \times & \left[\int_{-1/v_{{\bar{\alpha}\bar{\nu}}}}^{+1/v_{{\bar{\alpha}\bar{\nu}}}}
dz\,F_0 (z)\right]\nonumber \\
& \times & \Bigl({\Omega\over 4\pi\sqrt{v_{c0}\,v_{s1}}}
\Bigr)^{\zeta_0 (q',\,q'')}\left({2l[\omega_{BL} (k)-\omega]\over
v_{c0}\,v_{s1}}\right)^{-{1\over 2}} \, , \label{BLba}
\end{eqnarray}
where $a_{\alpha\nu} (q)=\partial v_{\alpha\nu} (q)/\partial q$.
In turn, just above ($l=+1$) or below ($l=-1$) the line the weight
distribution reads,
\begin{eqnarray}
& & B_{{\cal{N}}}^{l} (k,\,\omega)  \approx {2\,\Theta
\Bigl(\Omega -l[\omega -\omega_{BL} (k)]\Bigr)\over
\pi\,C_c\,C_s\,\zeta_0 (q',\,q'')\,\sqrt{\vert a_{\alpha'\nu'}
(q')\vert +\vert a_{\alpha''\nu''}
(q'')\vert}}
\nonumber \\
& \times & \Bigl\{\Theta\Bigl(v_{{\bar{\alpha}\bar{\nu}}}\,[1-{l[\omega -
\omega_{BL}(k)]\over\Omega}]-\vert v_{\alpha'\nu'}
(q')\vert\Bigr)\int_{-1\over v_{{\bar{\alpha}\bar{\nu}}}}^{1\over v_{{\bar{\alpha}\bar{\nu}}}}dz\,F_0 (z)
\nonumber \\
& + & {\rm sgn} (q')\,\theta\Bigl(\vert v_{\alpha'\nu'}
(q')\vert-v_{{\bar{\alpha}\bar{\nu}}}\,[1-{l[\omega - \omega_{BL}(k)]\over\Omega}]\Bigr)
\nonumber \\
& \times & 
\int_{-{{\rm sgn} (q')\,1\over v_{{\bar{\alpha}\bar{\nu}}}}}^{{1\over v_{\alpha'\nu'}
(q')}(1-{l[\omega-\omega_{BL}(k)]\over\Omega})}dz\,F_0 (z)\Bigr\}
\nonumber \\
& \times & \Bigl[\Bigl({\Omega\over 4\pi\sqrt{v_{c0}\,v_{s1}}}
\Bigr)^{\zeta_0 (q',\,q'')}-\Bigl({l[\omega -\omega_{BL} (k)]\over
4\pi\sqrt{v_{c0}\,v_{s1}}\,[1 -z\,v_{\alpha'\nu'}(q')]}
\Bigr)^{\zeta_0 (q',\,q'')}\Bigr]
\nonumber \\
& \times & \left({2l[\omega -\omega_{BL}
(k)]\over v_{c0}\,v_{s1}}\right)^{-{1\over 2}} \, , \label{BLab}
\end{eqnarray}
where $\theta (x)=0$ for $x\leq 0$ and $\theta (x)=1$ for
$x>0$, $\Omega$ is the energy cut off of the elementary
processes (C) defined in Ref. \cite{V-1}, $\zeta_0=\zeta_0 (q',\,q'')$
is the functional given in Eq. (\ref{zeta0}) for the excited states 
which contribute to the weight distribution (66) of that reference,
and $C_c$ and $C_s$ are defined in Eq. (68) of Ref. \cite{V}. 
\begin{figure}
\includegraphics[scale=.8]{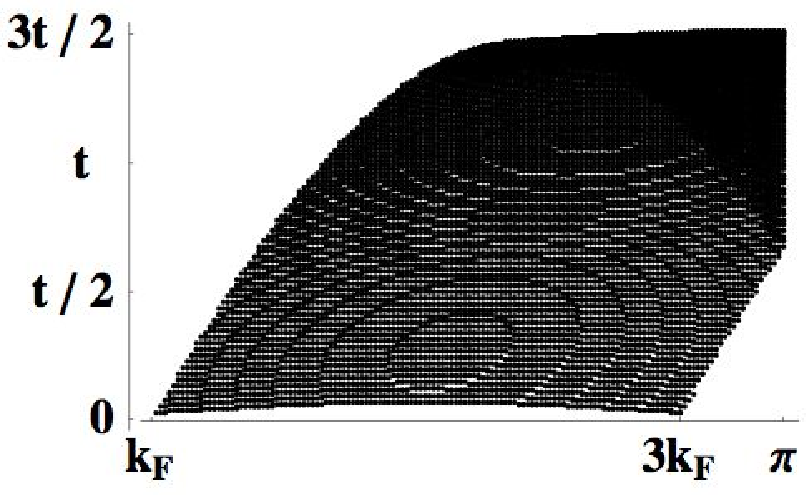} 
\includegraphics[scale=0.8]{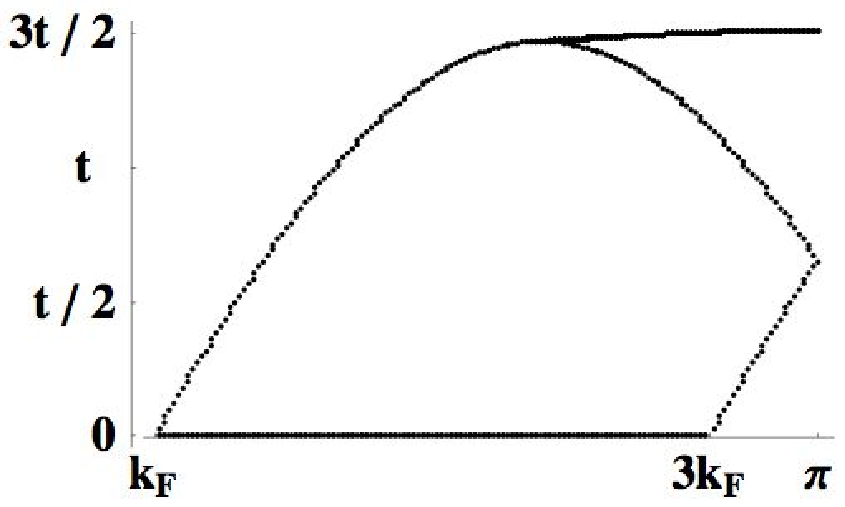} 
\caption{\label{fig8}
The region of the $(k,\omega)$ plane with a finite one-electron
addition spectral weight obtained by running over all the momentum 
values of the $c0$ and $s1$ momentum distribution
deviations of Eqs. (\ref{DNq-ER1s1H}) and (\ref{DNqm0-ER1s1H})
 for $U/t=100$, $n=0.59$, and $m\rightarrow 0$.
The corresponding branch and border lines are also shown.}
\end{figure}

\section{APPLICATIONS OF THE PSEUDOFERMION DYNAMICAL THEORY TO THE
SPECTRUM OF TTF-TCNQ}
\label{SecIV}

As discussed in Ref. \cite{TTF}, the spectral-weight distribution of TTF-TCNQ is fully determined by
the occupancy configurations of the $c$ and $s1$ pseudofermions in the one-electron
excited states. The studies of that reference reveal that for 
electronic density $n=1.41$ the electron removal spectrum calculated for $t=0.35$ eV and 
$U=1.96$ eV ($U/t=5.61$) yields the best agreement with the TTF related experimental dispersions. 
In turn, for electronic density $n=0.59$ an almost perfect agreement with the TCNQ 
related experimental dispersions is reached for the finite-energy-electron-removal spectrum
calculated for $t = 0.40$ eV and $U=1.96$ eV ($U/t=4.90$) \cite{TTF,spectral}. 
 
If one profits from the model particle-hole symmetry, one can achieve the one-electron removal
spectrum at $n=1.41$ and a given $U/t$ value from the one-electron addition spectrum
at $n=0.59$ and the same value of $U/t$. The full theoretical line shape of the one-electron 
removal spectral-weight distribution found in Ref. \cite{TTF}
to fit the corresponding spectral features of TTF-TCNQ is shown in Fig. 1.
Note that here such a spectrum is rotated relative to that shown in 
Fig. 2 of that reference.

Let us consider the processes that contribute to the spectrum of Fig. 1. 
Each of the corresponding spectral-weight contributions to the full one-electron 
spectrum plotted in that figure are obtained by use of the 
PDT expressions provided in the previous section and Ref. \cite{V-1}.
In addition to the values of $U/t$ and $n$ specific to TTF-TCNQ, some of the 
spectral contributions plotted below refer to variations in these parameters.
That allows us to study how the spectral features evolve under such
variations. The study of the spectral-weight peaces contributing to the full spectrum plotted
in Fig. 1 allows the identification of the contributions from the
different expressions considered
in the previous section and Ref. \cite{V-1}, including the border line expressions  
derived in this paper and provided in Eqs. (\ref{BLba}) and (\ref{BLab}).
\begin{figure}
\includegraphics[scale=.8]{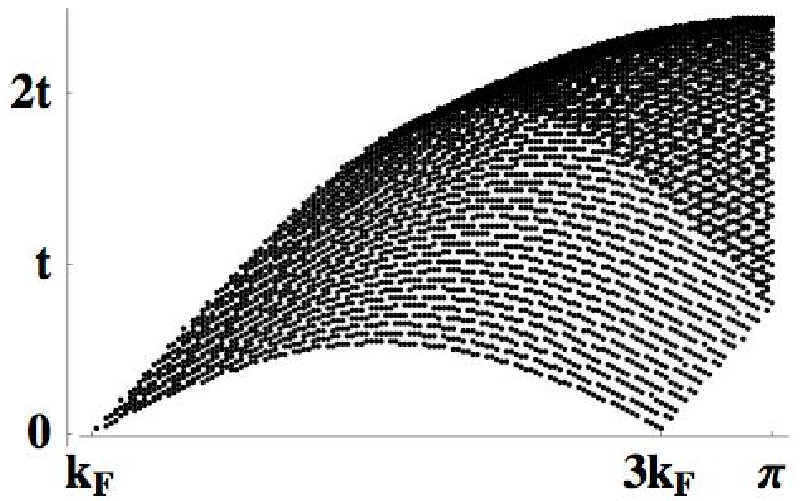} 
\includegraphics[scale=0.8]{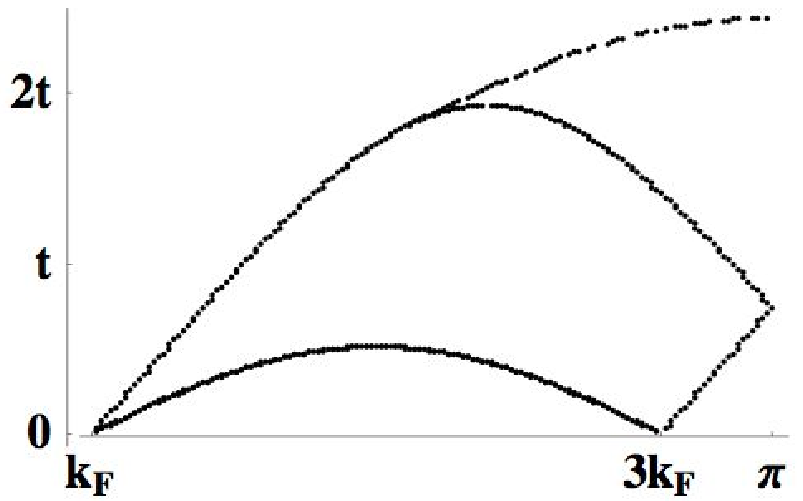} 
\caption{\label{fig9}
The same region of the $(k,\omega)$ plane and branch and border
lines as in Fig. 8 for $U/t=5.61$, $n=0.59$, and $m\rightarrow 0$.}
\end{figure}

The one-electron removal spectral-weight distribution of Fig. 1 has two main contributions
referring to the TCNQ and TTF related spectral features, respectively.
The TCNQ related spectrum is obtained by considering the one-electron
removal spectral function for $n=0.59;t=0.40\hspace{0.1cm}{\rm eV};U/t=4.90$.
In turn, we profit from the model particle-hole symmetry to derive the TTF 
related photoemission spectrum for $n=1.41;t = 0.35\hspace{0.1cm}{\rm eV};U/t=5.61$
from the corresponding one-electron addition spectrum for
$n=0.59;t = 0.35\hspace{0.1cm}{\rm eV};U/t=5.61$ and changing the
energy and momentum signs.

Nearly the whole one-electron removal spectral weight corresponds
to excitations described by the following deviations from the
ground-state
$N_{\alpha\nu}$ $\alpha\nu$ pseudofermion numbers and 
$N_{\alpha\nu}^h$ $\alpha\nu$ pseudofermion-hole numbers
where $\alpha\nu=c0,s1$,
\begin{equation}
\Delta N_{c0} = - \Delta N_{c0}^h = -1 \, ; \hspace{1cm} \Delta
N_{s1} = - \Delta N_{s1}^h = -1 \, , \label{DN-ER1s1H}
\end{equation}
and with the numbers $Q_{\alpha\nu}^0/2$ appearing
on the right-hand side of Eq. (\ref{Qcan1j}) reading,
\begin{equation}
Q_{c0}^0/2 = \pm\pi/2 \, ; \hspace{1cm} Q_{s1}^0/2 = 0 \, .
\label{pi-ER1s1H}
\end{equation}
If we consider bare-momentum continuum values, the general
deviations read,
\begin{eqnarray}
\Delta N_{c0} (q) & = & -{2\pi\over L}\,\delta (q -q_1) -{\pi\over
L}\,\delta (q \mp 2k_F) 
\nonumber \\
& + & {\pi\over L}\,\delta (q \pm 2k_F) \, ;
\hspace{0.25cm} q_1\in [-2k_F,\,+2k_F] \nonumber \\
\Delta N_{s1} (q) & = & -{2\pi\over L}\,\delta (q -{q'}_1) \, ;
\hspace{0.25cm} {q'}_1\in [-k_{F\downarrow},\,+k_{F\downarrow}] \,
, \label{DNq-ER1s1H}
\end{eqnarray}
and thus
\begin{equation}
\Delta N_{s1} (q) = -{2\pi\over L}\,\delta (q -{q'}_1) \, ;
\hspace{0.25cm} {q'}_1\in [-k_{F},\,+k_{F}] \label{DNqm0-ER1s1H}
\end{equation}
as $m\rightarrow 0$ for the initial ground state. Such deviations
and numbers (\ref{pi-ER1s1H}) are then used in the
general functional $Q_{\alpha\nu}(q_j)/2$ given in
Eq. (\ref{Qcan1j}), whose functional $Q^{\Phi}_{\alpha\nu} (q_j)/2$ 
is defined in Eq. (\ref{qcan1j}). Such a procedure leads to,
\begin{eqnarray}
Q_{c0}(q)/2 & = & \pm\pi/2 - \pi\Phi_{c0,\,c0}(q,q_1)  -
\pi\Phi_{c0,\,c0}(q,\mp 2k_F)/2 
\nonumber \\
& + & \pi\Phi_{c0,\,c0}(q,\pm 2k_F)/2 -
\pi\Phi_{c0,\,s1}(q,{q'}_1) \, , \label{Qc0-ER1s1H}
\end{eqnarray}
and
\begin{eqnarray}
Q_{s1}(q)/2 & = & - \pi\Phi_{s1,\,c0}(q,q_1)  - \pi\Phi_{s1,\,c0}(q,\mp
2k_F)/2 
\nonumber \\
& + & \pi\Phi_{s1,\,c0}(q,\pm 2k_F)/2 -
\pi\Phi_{s1,\,s1}(q,{q'}_1) \, , \label{Qs1-ER1s1H}
\end{eqnarray}
respectively. 
\begin{figure}
\includegraphics[scale=.8]{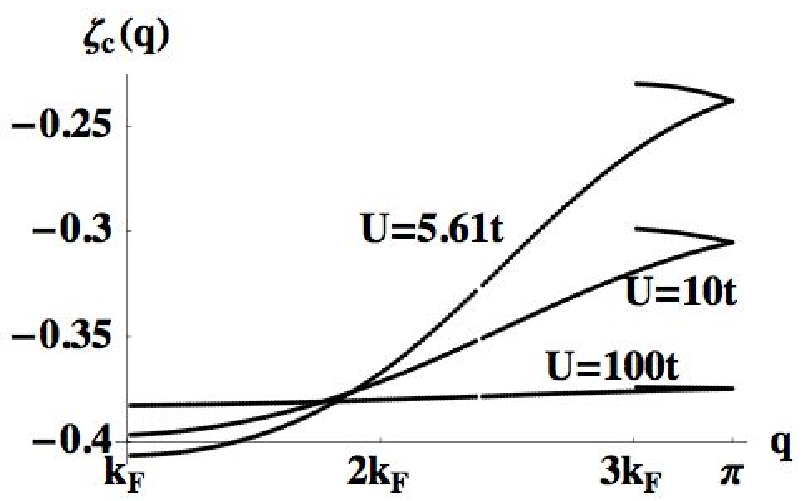} 
\includegraphics[scale=0.8]{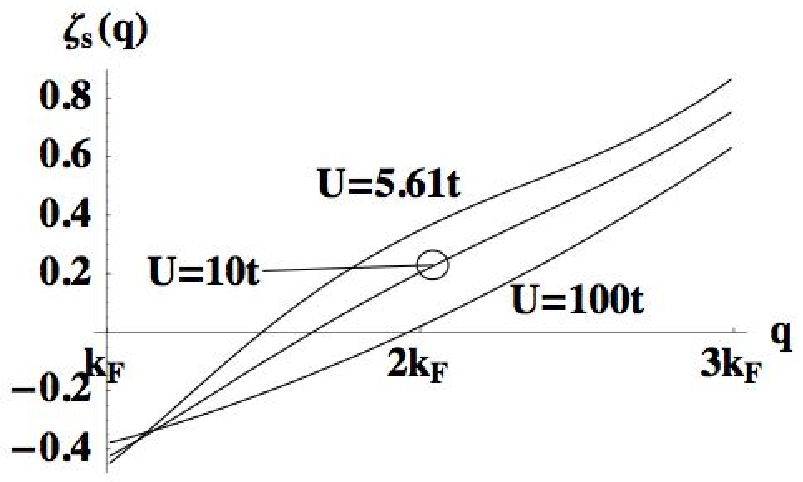} 
\caption{\label{fig10}
The value of the $c0$ and $s1$ branch lines 
exponents of the one-electron addition spectral-weight
distribution as a function of momentum for 
$n=0.59$, $m\rightarrow 0$, and several magnitudes
of $U/t$. The exponents of index $c$ and $s$ refer to
the $c0$ and $s1$ branch lines, respectively.}
\end{figure}

The creation of the holes in the $c0$ and $s1$ bands whose energy
dispersions are plotted in Figs. 6 and 7 of Ref. \cite{II-04},
respectively, corresponds for $U/t=100$, $n=0.59$, and
$m\rightarrow 0$ to a finite spectral-weight distribution in
the region of the $(k,\omega)$ plane shown in Fig. 2. 
The lines associated with the branch lines and border
lines in the finite-weight distribution are also shown.
Most lines of the one-electron spectral-weight distribution 
plotted in Fig. 1
corresponding to power-law singularities are of the
branch line type. In contrast to the border-line singularities
of expressions (\ref{BLba}) and (\ref{BLab}), whose
exponent $-1/2$ is momentum and $U/t$ independent,
that controlling the branch-line power-law singularities
is both momentum and interaction dependent. 
Near a branch line the general expression of the weight distribution 
is that provided in Eq. (70) of Ref. \cite{V-1}. 
Only for momentum and interaction values where the 
exponent of that expression is negative is
the corresponding line called a branch line.
For the limit $m\rightarrow 0$ 
considered here the pre-factor function $F_0 (z)$ appearing in
such a branch-line expression is that given
in Eqs. (\ref{C-v})-(\ref{C-v-3D0}) of Appendix A.
The expressions provided in the latter equations are a generalization of
the pre-factor function $F_0 (z)$ of Eq. (62) of Ref. \cite{V-1}.
The latter applies solely when $0<n<1$ and $0<m<n$. 

In figure 3 the same finite one-electron removal 
spectral-weight region and corresponding lines are shown for 
the $U/t=4.9$, $n=0.59$, and
$m\rightarrow 0$ values suitable to the TCNQ
related spectral features. Consistently with the $U/t$
dependence of the energy bandwidth of the
$s1$ pseudofermion dispersion plotted in Fig. 
7 of Ref. \cite{II-04}, note that the border line
connecting the minimum energy points of the
two $c0$-branch lines has
$v_{c0} (q)=v_{s1} (q')\approx 0$. Upon decreasing
the $U/t$ value to that $U/t=4.9$ suitable to the
TCNQ related spectral features of Fig. 1, such a border
line acquires a curvature. The curvature achieved
at $U/t=4.9$ is that behind
the agreement found in Ref. \cite{TTF} between
the PDT predictions and the photoemission spectral
features of TTF-TCNQ. That agreement is not obtained
for larger or smaller values of $U/t$.
\begin{figure}
\includegraphics[scale=.08]{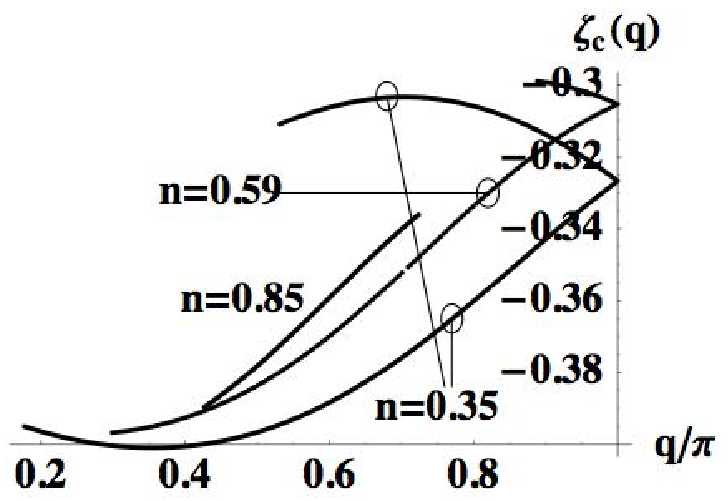} 
\includegraphics[scale=0.8]{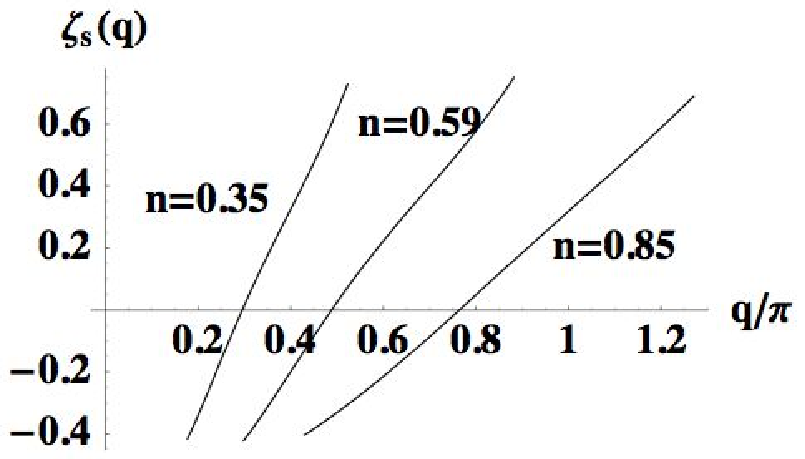} 
\caption{\label{fig11}
The same exponents as in Fig. 10 as a function of momentum for 
$U/t=10$, $m\rightarrow 0$, and several values of $n$.}
\end{figure}

The (negative-exponent) $c0$ branch line of the weight distributions
of Figs. 2 and 3 runs between the
excitation momenta $-k_F$ and $3k_F$
and the $s1$ branch line from
$-k_F$ to $k_F$. The corresponding
power-law momentum dependent exponents are
plotted in Fig. 4 for $n=0.59$, $m\rightarrow 0$,
and several values of $U/t$. The $c0$ branch line
segment between $-k_F$ and $0$ is folded
in the positive momentum region. While the 
negative $c0$ branch line exponent is smaller 
for $U/t$ large, the negative $s1$ branch line 
exponent is smaller for smaller values of
$U/t$. In order to illustrate how systematic variations
in the parameters lead to the evolution of the
spectral features, the same exponents as in Fig. 4 
are plotted in Fig. 5 as a function of momentum for 
$U/t=10$, $m\rightarrow 0$, and several values of $n$.

The use in the general PDT expressions provided in this paper
and in Ref. \cite{V-1} of the specific $c0$
and $s1$ momentum deviations of Eqs. (\ref{DNq-ER1s1H}) and (\ref{DNqm0-ER1s1H}) 
and corresponding functionals given in Eqs. (\ref{Qc0-ER1s1H}) and (\ref{Qs1-ER1s1H})
leads for $U/t=100$, $n=0.59$, and $m\rightarrow 0$ to the
one-electron removal spectral-weight distribution plotted in Fig. 6. 
Such a large-$U/t$ distribution is quite similar to the function
$B (k,\omega)$ plotted in Fig. 1 of Ref. \cite{Penc} for
$U/t\rightarrow\infty$. However, the method used in the
studies of the latter reference does not apply to finite
values of $U/t$. The one-electron removal spectral-weight distribution 
related to the TCNQ spectral features is plotted in Fig. 7
and refers instead to $U/t=4.9$, $n=0.59$, and $m\rightarrow 0$.
It is a part of the full one-electron spectral weight distribution
plotted in Fig. 1. 

Most singular behaviours of the one-electron removal spectral-weight 
distributions of Figs. 6 and 7 refer to $c0$ and $s1$ branch lines whose
negative power-law exponents magnitude depends on $U/t$ and the momentum. 
These exponents are plotted in Figs. 4 and 5. In turn, the weaker
lines connecting in Figs. 2 and 3 the minimum energy points of the
two $c0$ lines are border lines. The singularities of the latter lines correspond
instead to the $U/t$ and momentum independent exponent $-1/2$,
as given in the general border-line expressions (\ref{BLba}) and (\ref{BLab})
derived in this paper.
\begin{figure}
\includegraphics[scale=0.8]{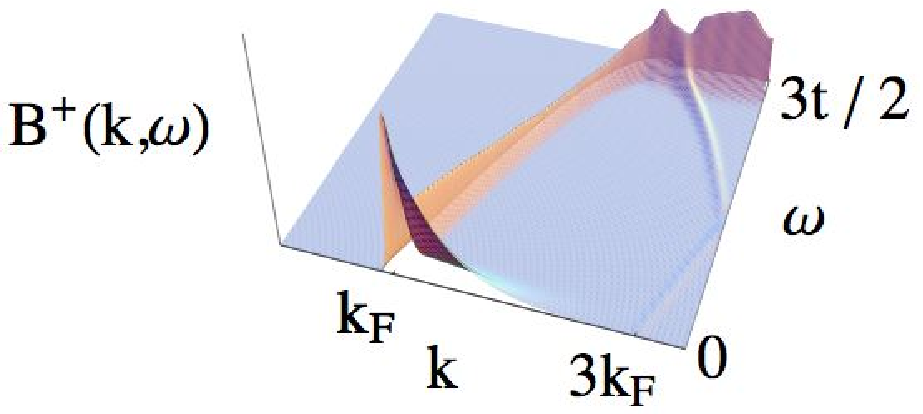} 
\includegraphics[scale=0.8]{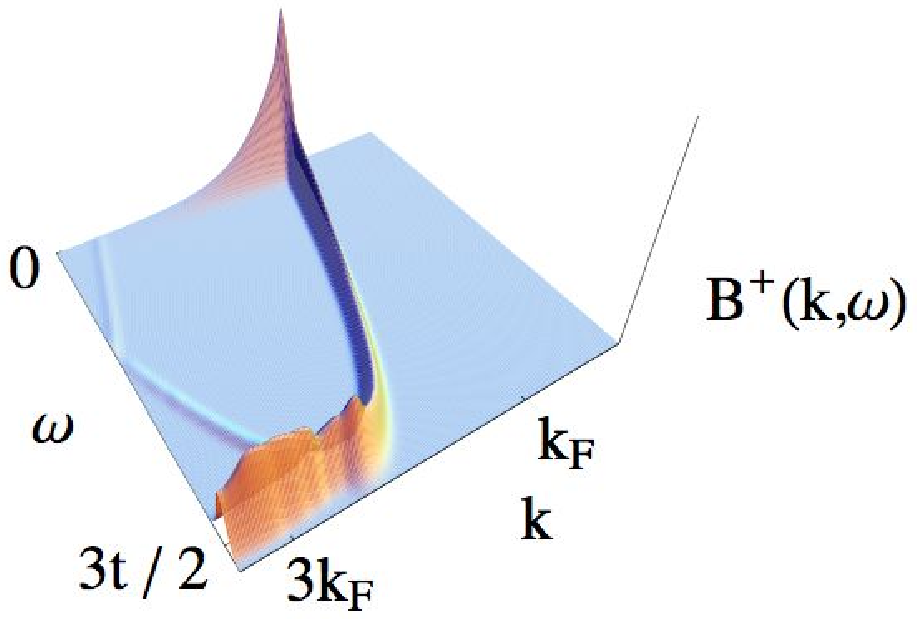} 
\caption{\label{fig12}
Full PDT one-electron addition spectral-weight distribution 
associated with the $c0$ and $s1$ momentum deviations of 
Eqs. (\ref{DNq-LEA1s1H}) and (\ref{DNqm0-LEA1s1H})
and corresponding functionals given in 
Eqs. (\ref{Qc0-LEA1s1H}) and (\ref{Qs1-LEA1s1H}) 
for $U/t=100$, $n=0.59$, and $m\rightarrow 0$.}
\end{figure}

Next let us consider the lower-Hubbard-band one-electron addition spectrum for
$n=0.59;t = 0.35\hspace{0.1cm}{\rm eV};U/t=5.61$ from which
one derives the TTF related photoemission one-electron removal spectrum for 
$n=1.41;t = 0.35\hspace{0.1cm}{\rm eV};U/t=5.61$.
Nearly the whole lower-Hubbard-band one-electron addition spectral weight corresponds
to the excitations leading to the following deviations from the ground-state
$N_{\alpha\nu}$ $\alpha\nu$ pseudofermion numbers and 
$N_{\alpha\nu}^h$ $\alpha\nu$ pseudofermion-hole numbers
where $\alpha\nu=c0,s1$,
\begin{equation}
\Delta N_{c0} = - \Delta N_{c0}^h = +1 \, ; \hspace{1cm} \Delta
N_{s1} = 0 \, ; \hspace{1cm} \Delta N_{s1}^h = +1 \, ,
\label{DN-LEA1s1H}
\end{equation}
and with the numbers $Q_{\alpha\nu}^0/2$ appearing
on the right-hand side of Eq. (\ref{Qcan1j}) reading,
\begin{equation}
Q_{c0}^0/2 = 0 \, ; \hspace{1cm} Q_{s1}^0/2 = \pm\pi/2 \, .
\label{pi-LEA1s1H}
\end{equation}
If we choose the
excitation branches that survive for $m\rightarrow 0$, the general
deviations are for bare-momentum continuum values given by,
\begin{eqnarray}
\Delta N_{c0} (q) & = & {2\pi\over L}\,\delta (q -q_1)  \, ;
\hspace{0.25cm} q_1\in [-2k_F,\,+\pi] \hspace{0.3cm}{\rm
and}\hspace{0.3cm}
q_1\in [-\pi,\,+2k_F] \nonumber \\
\Delta N_{s1} (q) & = & -{2\pi\over L}\,\delta (q -{q'}_1) +
{\pi\over L}\,\delta (q +k_{F\downarrow}) 
\nonumber \\
& + & {\pi\over L}\,\delta
(q -k_{F\downarrow}) \, ; \hspace{0.25cm} {q'}_1\in
[-k_{F\downarrow},\,+k_{F\downarrow}] \, , \label{DNq-LEA1s1H}
\end{eqnarray}
and thus
\begin{eqnarray}
\Delta N_{s1} (q) & = & -{2\pi\over L}\,\delta (q -{q'}_1)  +
{\pi\over L}\,\delta (q -k_{F}) 
\nonumber \\
& + & {\pi\over L}\,\delta (q
+k_{F})\, ; \hspace{0.25cm} {q'}_1\in [-k_{F},\,+k_{F}]
\label{DNqm0-LEA1s1H}
\end{eqnarray}
as $m\rightarrow 0$ for the initial ground state.
Alike for the above one-electron removal case, such deviations
and the numbers (\ref{pi-LEA1s1H}) are used in the
general functional $Q_{\alpha\nu}(q_j)/2$ provided in
Eq. (\ref{Qcan1j}) and its functional $Q^{\Phi}_{\alpha\nu} (q_j)/2$ 
defined in Eq. (\ref{qcan1j}). One then finds,
\begin{eqnarray}
Q_{c0}(q)/2 & = & \pi\Phi_{c0,\,c0}(q,q_1) -
\pi\Phi_{c0,\,s1}(q,{q'}_1) 
\nonumber \\
& + & \pi\Phi_{c0,\,s1}(q,-k_{F\downarrow})/2 +
\pi\Phi_{c0,\,s1}(q,+k_{F\downarrow})/2 \, , \label{Qc0-LEA1s1H}
\end{eqnarray}
and
\begin{eqnarray}
Q_{s1}(q)/2 & = & \pm\pi/2 + \pi\Phi_{s1,\,c0}(q,q_1) -
\pi\Phi_{s1,\,s1}(q,{q'}_1) 
\nonumber \\
& + & \pi\Phi_{s1,\,s1}(q,-k_{F\downarrow})/2 +
\pi\Phi_{s1,\,s1}(q,+k_{F\downarrow})/2 \, , \label{Qs1-LEA1s1H}
\end{eqnarray}
respectively.
\begin{figure}
\includegraphics[scale=0.8]{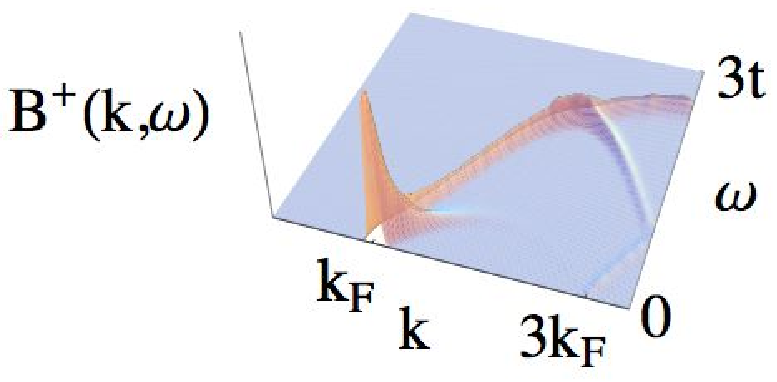} 
\includegraphics[scale=0.8]{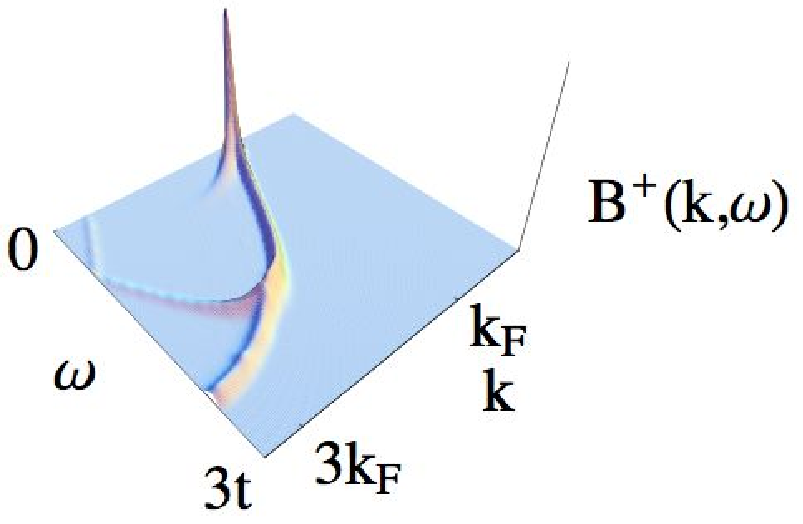} 
\caption{\label{fig13}
The same PDT one-electron addition spectral-weight distribution 
as in Fig. 8 for the values $U/t=5.61$, $n=0.59$, and 
$m\rightarrow 0$ suitable to the TTF related spectral
features.}
\end{figure}

The creation of the $c0$ pseudofermion and 
$s1$ pseudofermion hole refers for $U/t=100$, $n=0.59$, and
$m\rightarrow 0$ to a finite spectral-weight distribution in
the region of the $(k,\omega)$ plane shown in Fig. 8. 
The corresponding lines associated with potential 
branch lines and border lines are also plotted. In Fig.
9 the same finite spectral-weight
region is shown for the $U/t=5.61$, $n=0.59$, and
$m\rightarrow 0$ values suitable to the TTF
related spectral features. The border line
connecting the maximum energy point of the
$c0$-branch line at momentum $2k_F$
with the maximum energy point of the finite
spectral-weight distribution at momentum
$\pi$ has in Fig. 8 $v_{c0} (q)=v_{s1} (q')\approx 0$. Upon decreasing
$U/t$ to the value $U/t=5.61$ suitable to the
TTF related spectral features of Fig. 1 that border
line acquires a curvature, as shown in Fig. 9. Such a curvature 
and corresponding value of $U/t$ are those
behind the agreement achieved in Ref. \cite{TTF} between
the PDT predictions and the photoemission spectral
features of TTF-TCNQ.

The $c0$ branch lines of the weight distributions
of Figs. 8 and 9 run between the
excitation momenta $k_F$ and $\pi$
and between $3k_F$ and $\pi$, respectively.
In turn, the maximum extension of the $s1$ branch line 
is achieved for large values of $U/t$ when it runs from
$k_F$ to near $3k_F$. The corresponding
power-law momentum dependent exponents are
plotted in Fig. 10 for $n=0.59$, $m\rightarrow 0$,
and several values of $U/t$. While the 
$c0$ branch lines exponents are negative,
the $s1$ branch line exponent is negative for
momenta smaller than approximately $k_F$ and
$2k_F$ for $U/t=5.61$ and $U/t=100$, respectively.
Again in order to illustrate the evolution of the
spectral features upon varying the parameters, 
the same exponents as in Fig. 10
are plotted in Fig. 11 as a function of momentum for 
$U/t=10$, $m\rightarrow 0$, and several values of $n$.

The general PDT spectral-function expressions 
provided in this paper and in Ref. \cite{V-1} lead for 
$U/t=100$, $n=0.59$, and $m\rightarrow 0$ and the specific $c0$
and $s1$ momentum deviations of Eqs. (\ref{DNq-LEA1s1H}) and (\ref{DNqm0-LEA1s1H})
and corresponding functionals provided in 
Eqs. (\ref{Qc0-LEA1s1H}) and (\ref{Qs1-LEA1s1H}) to the
one-electron addition spectral-weight distribution plotted in Fig. 12. 
The large-$U/t$ distribution plotted in that figure is quite similar to 
the function $A (k,\omega)$ plotted in Fig. 1 of Ref. \cite{Penc} for
$U/t\rightarrow\infty$.
The one-electron addition spectral-weight distribution 
related to the TTF spectral features is plotted in Fig. 13
and refers instead to $U/t=5.61$, $n=0.59$, and $m\rightarrow 0$.
After a straightforward particle-hole transformation, the latter distribution
leads to the one-electron removal spectral-weight distribution 
suitable to the TTF related spectral features. The latter corresponds
to $U/t=5.61$, $n=1.41$, and $m\rightarrow 0$ and
is part of the full one-electron removal spectral-weight distribution
plotted in Fig. 1.

Most singular behaviours of the one-electron addition spectral-weight 
distributions of Figs. 12 and 13 correspond to $c0$ and $s1$ branch lines whose
negative power-law exponents are dependent on the $U/t$ and momentum values. 
These exponents are plotted in Figs. 10 and 11 and the power-law singularities
occur for the momenta where they are negative. In turn, the 
line connecting in Figs. 8 and 9 the maximum energy point of the
$c0$-branch line at momentum $2k_F$
with the maximum energy point of the finite
spectral-weight distribution at momentum $\pi$
is a border line. The singularities at the latter line correspond
to the $U/t$ and momentum independent exponent $-1/2$
and near it the line shape is of the general
form provided in expressions (\ref{BLba}) and (\ref{BLab}).

The four functionals $2\Delta_{\alpha\nu}^{\iota}$
where $\alpha\nu=c0,s1$ and $\iota=\pm 1$ given in Eq.
(\ref{zeta0}) are in the case of the one-electron excitations
behind the spectral-weight distribution plotted in Fig. 1
fully controlled by the functionals provided in
Eqs. (\ref{Qc0-ER1s1H}) and (\ref{Qs1-ER1s1H})
for $U/t=4.90$, $n=0.59$, and $m\rightarrow 0$
for the TCNQ related spectral features and in
Eqs. (\ref{Qc0-LEA1s1H}) and (\ref{Qs1-LEA1s1H}) 
for $U/t=5.61$, $n=0.59$, and $m\rightarrow 0$
for the TTF related spectral features. Since the
magnitude of the latter functionals is a function
of two momenta $q_1$ and ${q'}_1$ of the
deviations of Eqs. (\ref{DNq-ER1s1H}) and (\ref{DNqm0-ER1s1H}) 
or Eqs. (\ref{DNq-LEA1s1H}) and (\ref{DNqm0-LEA1s1H})
belonging to the $c0$ and $s1$
band, respectively, it is different for each point 
of the $(k,\omega)$ plane. Moreover, for some of
the latter points there are contributions from more
than one pair of momenta of $q_1$ and ${q'}_1$. 

We emphasize that depending on whether none,
one, or two of the two above momenta 
$q_1$ and ${q'}_1$ is or are Fermi points, the 
corresponding functionals $2\Delta_{\alpha\nu}^{\iota}$
have different form and the excitation contributes to
line shapes described by the expressions
provided in Eqs. (66), (70), and (68) of Ref.
\cite{V-1}, respectively. All such expressions involve
the pre-factor function $F_0 (z)$ given in Eqs. (\ref{C-v})-(\ref{C-v-3D0}) of Appendix A.
The expressions provided in these equations are a generalization of
the pre-factor function $F_0 (z)$ provided in Eq. (62) of Ref. \cite{V-1}.
The latter applies when the four functionals $2\Delta_{\alpha\nu}^{\iota}$
where $\alpha\nu=c0,s1$ and $\iota=\pm 1$ are finite. 
However, in the limit $m\rightarrow 0$ that the 
spectral-weight distributions studied in this paper refer to one must use in
the general expressions provided in Eqs. (66), (70), and (68) of Ref.
\cite{V-1} the pre-factor function $F_0 (z)$ of 
Eqs. (\ref{C-v})-(\ref{C-v-3D0}) of Appendix A.

Note that the above mentioned general expressions 
provided in Eqs. (68) and (70) of Ref. \cite{V-1} describe the
line shape near well-define points and lines in the
$(k,\omega)$ plane, respectively. When the corresponding exponents
are negative, such spectral features refer to the
point and branch-line singularities of the weight
distributions plotted in Figs. 1, 6, 7, 12, and 13.
In turn, the only type of singularity occurring within
the line shape described by the above mentioned general 
expression provided in Eq. (66) of Ref. \cite{V-1} is
that associated with the border lines studied in
the previous section. The corresponding expressions 
(\ref{BLba}) and (\ref{BLab}) refer to the vicinity of such border lines.
\begin{figure}
\includegraphics[scale=0.4]{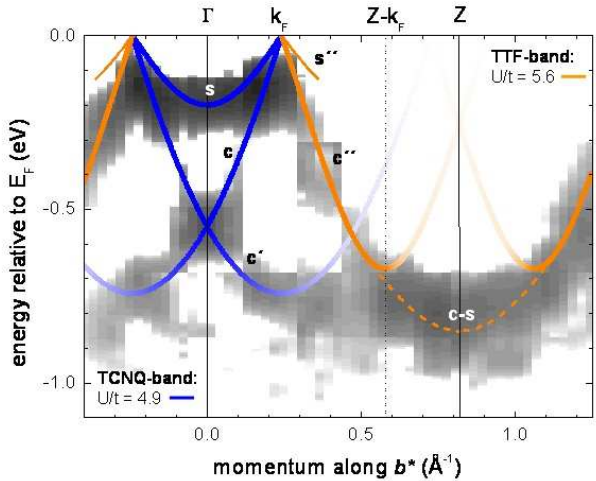} \caption{\label{fig14}
Experimental peak dispersions (grey scale) obtained by ARPES on TTF-TCNQ 
along the easy-transport axis as given in Fig. 7 of Ref. \cite{spectral0}
and matching theoretical branch and border lines. (The Z-point corresponds 
to the momentum $k=\pi$.) The corresponding detailed theoretical 
spectral-weight distributions over the whole $(k,\,\omega)$-plane are 
plotted above in Fig. 1. While the
theoretical charge-$c''$ and spin-$s''$ branch lines and $c-s$ 
border line refer upon the particle-hole transformation to the TTF related 
spectral features of Figs. 9 and 13, the charge-$c$, spin-$s$, and charge-$c'$ branch lines
correspond to the TCNQ related dispersions of Figs. 3 and 7.}
\end{figure}

The complementary studies of Ref. \cite{spectral} provide the
specific expressions of the four functionals $2\Delta_{\alpha\nu}^{\iota}$
where $\alpha\nu=c0,s1$ and $\iota=\pm 1$ given in Eq.
(\ref{zeta0}) for each of the point and branch-line singularities
of the TCNQ related one-electron removal spectral features.
The functionals given in Eqs. (\ref{Qc0-LEA1s1H}) and (\ref{Qs1-LEA1s1H}) 
are the basis of the derivation of similar expressions for
the TTF related spectral features. Those are straightforwardly
derived by choosing particular pairs of momentum values 
$q_1$ and ${q'}_1$ in the 
$c0$ and $s1$ momentum deviations of 
Eqs. (\ref{DNq-LEA1s1H}) and (\ref{DNqm0-LEA1s1H}), respectively. 
As discussed above, when such choices involve none,
one, or both such momentum pairs being a $c0$ or $s1$
Fermi point, the corresponding four functionals $2\Delta_{\alpha\nu}^{\iota}$
have different forms and the excitation contributes to
line shapes described by the expressions
provided in Eqs. (66), (70), and (68) of Ref.
\cite{V-1} for the two-dimensional weight distribution, 
vicinity of branch lines, and vicinity of
singular points, respectively. 

Finally, in Fig. 14 we plot the theoretical lines corresponding to
the sharpest spectral 
features considered in Fig. 1 but omit the corresponding detailed spectral-weight distribution
over the $(k,\,\omega)$-plane provided in that figure. 
The figure also displays the experimental dispersions in the electron removal 
spectrum of TTF-TCNQ as measured by ARPES in Ref. \cite{spectral0}.
The border line connecting the maximum energy point of the
$c0$-branch line at momentum $2k_F$
with the maximum energy point of the finite
spectral-weight distribution at momentum
$\pi$ in Fig. 9 leads upon the particle-hole transformation
to that called $c-s$ line in Fig. 14. In turn the border line
connecting the minimum energy points of the
two $c0$ lines of Fig. 3 associated with
the TCNQ related spectral features is not marked
in Fig. 14, yet the corresponding experimental
spectral weight is clearly visible.   

\section{DISCUSSION AND CONCLUDING REMARKS}
\label{SecV}

In this paper we have generalized the closed-form analytical
expressions for the finite-energy one- and two-electron
spectral-weight distributions of a 1D correlated metal with
on-site electronic repulsion introduced in Ref. \cite{V-1} to
all electronic densities of the metallic phase and
zero spin density. Moreover, we have studied the
particular form of the expressions derived here for the
processes contributing to the 
one-electron spectral-weight distributions related to the TTF and TCNQ 
stacks of molecules, respectively, in the quasi-1D organic compound TTF-TCNQ
investigated in Ref. \cite{TTF}. The corresponding
full theoretical one-electron spectral-weight distribution plotted
in Fig. 1 agrees quantitatively for the whole experimental energy 
bandwidth with the observed one-electron spectral features shown in Fig. 14.

Other applications of our finite-energy spectral-weight-distribution expressions 
to several materials, correlated quantum systems, and spectral functions are in
progress. This includes the use of our theoretical results in the
two-atom spectral-weight distributions measured in 1D optical
lattices. Such studies also involve the Hubbard model
but with the electrons replaced by fermionic spin-$1/2$ atoms. While for the
one-electron features the branch lines play the major role, in the
case of two-electron spectral functions such as the charge dynamical
structure factor the two-pseudofermion spectral features of the form
given in Eq. (66) of Ref. \cite{V-1} lead to the main contributions.
Generalization of that equation to all electronic densities of the
metallic phase and to zero spin density involves the use of the
following expression,
\begin{eqnarray}
B_{{\cal{N}}}^{l} (k,\,\omega) & \approx & {1\over
\pi\,C_c\,C_s}
\left[\int_{-1/v_{{\bar{\alpha}\bar{\nu}}}}^{+1/v_{{\bar{\alpha}\bar{\nu}}}} dz\,F_0
(z)\right]\,\Bigl({\Omega\over 4\pi\sqrt{v_{c0}\,v_{s1}}}
\Bigr)^{\zeta_0 (q',\,q'')}
\nonumber \\
& \times & {\sqrt{v_{c0}\,v_{s1}}\over \zeta_0
(q',\,q'')\,\vert v_{\alpha'\nu'} (q')-v_{\alpha''\nu''} (q'')\vert}\, \,
; \hspace{0.25cm} l= \pm 1 \, , \label{B-2D-inte}
\end{eqnarray}
with the pre-factor function $F_0 (z)$ given by the expressions 
(\ref{C-v})-(\ref{C-v-3D0}) of Appendix A introduced in this paper. Alike in
Eq. (\ref{BLba}), here $q'$ and $q''$ stand for the bare-momentum values of the
created $\alpha'\nu'$ and $\alpha''\nu''$ pseudofermions or holes,
respectively. Indeed, the singular border-line function (\ref{BLba}) is a particular 
case of the general function provided in Eq. (\ref{B-2D-inte}), which
can be obtained by considering that $v_{\alpha'\nu'} (q')=v_{\alpha''\nu''} (q'')$ 
in the latter equation.

For finite values of $U/t$ the dominant term of the two-electron spectral functions
is often of the general form given in Eq. (\ref{B-2D-inte}). However,
it can also occur that for intermediate values of $U/t$ such functions 
are the sum of two or three 
dominant functions of the general form (\ref{B-2D-inte}).
The main contributions correspond to the two created 
objects being such that (i) $\alpha'\nu'=c0$ and 
$\alpha''\nu''=s1$,  (ii) $\alpha'\nu'=\alpha''\nu''=c0$, and (iii) 
$\alpha'\nu'=\alpha''\nu''=s1$. The relative importance of
the functions (i)-(iii) and whether the two created objects are
pseudofermions and/or pseudofermion holes depends on 
the specific two-electron spectral function under consideration. 
Our general expressions 
(\ref{C-v})-(\ref{C-v-3D0}) of Appendix A for the pre-factor function $F_0 (z)$
appearing on the right-hand side of Eq. (\ref{B-2D-inte}) 
allows the evaluation of all two-electron spectral-weight
distributions for initial ground states with 
zero spin density, which in many situations 
is the case of physical interest. The branch lines and
other spectral features also contribute to the two-electron
spectral-weight distributions, yet the dominant contributions
are of the general form given in Eq. (\ref{B-2D-inte}).
\begin{figure}
\includegraphics[scale=.4]{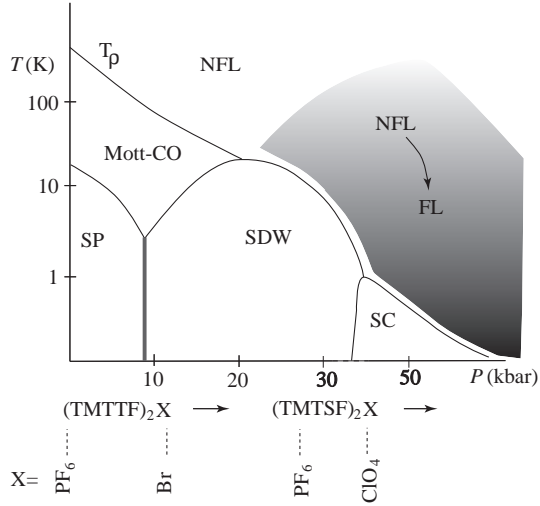} 
\caption{\label{fig15}
Temperature-pressure phase diagram of the (TMTTF)$_2$X and (TMTSF)$_2$X series of
compounds. It includes non-Fermi liquid (NPL), insulating Mott-Hubbard (Mott-CO),
spin-density-wave (SDW), spin-Peierls, Fermi-liquid (FL), and superconducting
(SC) phases.}
\end{figure}

For instance, the preliminary results of Ref. \cite{super} 
consider both one- and two-electron spectral features,
profits from the PDT for the one-chain problem, are consistent with the
phase diagram observed in the (TMTTF)$_2$X and (TMTSF)$_2$X series
of organic compounds, and explain the absence of superconductive
phases in TTF-TCNQ. The studies of that reference combine the 
PDT for several one- and two-electron spectral functions with a 
Renormalization Group analysis to study the instabilities
of a system of weakly coupled Hubbard chains. 
For low values of the onsite repulsion $U$ and
of the doping $\delta=(1-n)$, the leading instability is towards a
superconducting state. The process includes excited states above a
small correlation pseudogap. Similar features appear in extended
Hubbard models in the vicinity of commensurate fillings. The
theoretical predictions of such studies are consistent with the phase diagram
observed in the (TMTTF)$_2$X and (TMTSF)$_2$X series of organic
compounds represented in Fig. 15. From the results of Ref. \cite{super}
one can infer that the low-temperature phase of the
coupled-chain system will show long-range superconducting order. 
However, the precise nature of this phase, and the symmetry of the order
parameter is dependent on the arrangement of the chains within the
material. The investigations of Ref. \cite{super} use the exponents obtained from the 
PDT for the one-chain problem and confirm that in 1D 
non-Fermi liquids weak inter-chain
hopping can induce superconductivity. At low temperature these
materials show a spin-Peierls or spin-density-wave phase. Under pressure,
the (TMTSF)$_2$X compounds are driven to a superconducting phase, which is
removed again if one further increases the pressure.

In addition to the finite-energy spectral features of 
the organic compound TTF-TCNQ \cite{TTF} and the
preliminary studies of Ref. \cite{super} on
the (TMTTF)$_2$X and (TMTSF)$_2$X series
of organic compounds, the
general spectral-function expressions derived in this
paper will be used elsewhere in the study of specific
one- and two-electron spectral functions and its quantitative
application to the unusual spectral properties of other low-dimensional
materials and systems. While the studies of this paper considered 
the 1D Hubbard model, which describes successfully some of the 
exotic properties observed in low-dimensional materials 
\cite{spectral0,spectral,Eric,super}, 
our results are of general nature for many integrable interacting 
problems \cite{Voit} and therefore have wide applicability.

We thank Pedro D. Sacramento 
for stimulating discussions and the support of ESF Science 
Program INSTANS, European Union Contract 12881 (NEST), 
FCT grants SFRH/BD/6930/2001, 
POCTI/FIS/58133/2004, and PTDC/FIS/64926/2006 and
and OTKA grant T049607.

\appendix

\section{The pre-factor function $F_0 (z)$}

In this Appendix a set of alternative expressions for the pre-factor 
function $F_0 (z)$ on the right-hand side of Eq. (\ref{B-breve-asym})
derived by use of Eqs. (\ref{B-J-i-sum-GG}),  
(\ref{B-J-i-sum-GG-D0}), (\ref{B-J-i-sum-GG-DD0}), (\ref{B-l-i-breve}),
and (\ref{B-breve-asym}) according to the values of the two parameters $2\Delta_{\alpha\nu}^{\pm 1}$ 
and two parameters $2\Delta_{\bar{\alpha}\bar{\nu}}^{\pm 1}$ are given.

\begin{enumerate}
\item When the four parameters $2\Delta_{\alpha\nu}^{\pm 1}$ 
where $\alpha\nu =c0,s1$ and $\iota =\pm 1$ are finite we find,
\begin{eqnarray}
& & F_0 (z) = 2D_0\,\sqrt{{v_{{\bar{\alpha}\bar{\nu}}}\over
v_{\alpha\nu}}}\int_0^1dx\int_{-1}^{+1}dy
\nonumber \\
& \times & \prod_{\iota' =\pm
1}{\Theta\Bigl(1 -x+{\rm sgn} (z)\,\iota'\,\Bigl[v_{{\bar{\alpha}\bar{\nu}}}\vert
z\vert-{v_{{\bar{\alpha}\bar{\nu}}}\over v_{\alpha\nu}}\,y\Bigr]\Bigr)\over \Gamma
(2\Delta_{{\bar{\alpha}\bar{\nu}}}^{\iota'})}
\nonumber \\
& \times & {\Theta \Bigl(x+{\rm sgn}
(z)\,\iota'\,y\Bigr)
\over \Gamma (2\Delta_{\alpha\nu}^{\iota'})}\nonumber \\
& \times & \left(\sqrt{{v_{\alpha\nu}\over v_{{\bar{\alpha}\bar{\nu}}}}}\,\Bigl[1 -x+{\rm sgn}
(z)\,\iota'\,\Bigl[v_{{\bar{\alpha}\bar{\nu}}}\vert z\vert-{v_{{\bar{\alpha}\bar{\nu}}}\over
v_{\alpha\nu}}\,y\Bigr]\Bigr]\right)^{2\Delta_{{\bar{\alpha}\bar{\nu}}}^{\iota}-1}
\nonumber \\
& \times & \left(\sqrt{{v_{{\bar{\alpha}\bar{\nu}}}\over
v_{\alpha\nu}}}\,\Bigl[x+{\rm sgn}
(z)\,\iota'\,y\Bigr]\right)^{2\Delta_{\alpha\nu}^{\iota'}-1} \, .
\label{C-v}
\end{eqnarray}
\item When $2\Delta_{\bar{\alpha}\bar{\nu}}^{-\bar{\iota}}=0$ and the remaining three parameters  are
finite,
\begin{eqnarray}
& & F_0 (z) = 2D_0\,{v_{{\bar{\alpha}\bar{\nu}}}\over
v_{\alpha\nu}}\int_{-1}^{+1}dy
\nonumber \\
& \times & {\Theta\Bigl(\bar{\iota}\,[{\rm sgn} (z)y-v_{\alpha\nu}\,
\Bigl(z - {\bar{\iota}\over v_{{\bar{\alpha}\bar{\nu}}}}\Bigr)]\Bigr)
\Theta\Bigl(\bar{\iota}\,[z v_{\alpha\nu}
- {\rm sgn} (z)y]\Bigr)\over \Gamma (2\Delta_{{\bar{\alpha}\bar{\nu}}}^{\bar{\iota}})}\nonumber \\
& \times &
\left(\sqrt{{v_{\alpha\nu}\over v_{{\bar{\alpha}\bar{\nu}}}}}\,2\bar{\iota}\,\Bigl[z - {\rm sgn} (z) {y\over v_{\alpha\nu}}\Bigr]
v_{{\bar{\alpha}\bar{\nu}}}\right)^{2\Delta_{{\bar{\alpha}\bar{\nu}}}^{\bar{\iota}}-1}
\nonumber \\
& \times & \prod_{\iota' =\pm 1}{\Theta \Bigl(1+{\rm sgn}(z)\Bigl[\iota' +\bar{\iota}\,
{v_{{\bar{\alpha}\bar{\nu}}}\over v_{\alpha\nu}}\Bigr]y- \bar{\iota}\,v_{{\bar{\alpha}\bar{\nu}}}\,z\Bigr)
\over \Gamma (2\Delta_{\alpha\nu}^{\iota'})} \nonumber \\
& \times & \left(\sqrt{{v_{{\bar{\alpha}\bar{\nu}}}\over
v_{\alpha\nu}}}\,\Bigl[1+{\rm sgn}(z)\Bigl[\iota' +\bar{\iota}\,
{v_{{\bar{\alpha}\bar{\nu}}}\over v_{\alpha\nu}}\Bigr]y- \bar{\iota}\,v_{{\bar{\alpha}\bar{\nu}}}\,z
\Bigr]\right)^{2\Delta_{\alpha\nu}^{\iota'}-1} \, .
\label{C-v-1D0}
\end{eqnarray}
\item When $2\Delta_{\alpha\nu}^{-\iota}=0$ and the remaining three parameters are
finite,
\begin{eqnarray}
& & F_0 (z) = 2D_0\,\int_{-1}^{+1}dy\, {\Theta\Bigl(\iota\,{\rm sgn} (z)y\Bigr)\over
\Gamma (2\Delta_{\alpha\nu}^{\iota})}
\left(\sqrt{{v_{{\bar{\alpha}\bar{\nu}}}
\over v_{\alpha\nu}}}\,\iota\,{\rm sgn} (z)2y\right)^{2\Delta_{\alpha\nu}^{\iota}-1}\nonumber \\
& \times &
\prod_{\bar{\iota}' =\pm 1}{\Theta \Bigl(1-{\rm sgn}(z)\Bigl[\iota +\bar{\iota}'\,
{v_{{\bar{\alpha}\bar{\nu}}}\over v_{\alpha\nu}}\Bigr]y+\bar{\iota}'\,v_{{\bar{\alpha}\bar{\nu}}}\,z\Bigr)
\over \Gamma (2\Delta_{{\bar{\alpha}\bar{\nu}}}^{\bar{\iota}'})} \nonumber \\
& \times & 
\left(\sqrt{{v_{\alpha\nu}\over v_{{\bar{\alpha}\bar{\nu}}}}}\,\Bigl[1-{\rm sgn}(z)\Bigl[\iota +\bar{\iota}'\,
{v_{{\bar{\alpha}\bar{\nu}}}\over v_{\alpha\nu}}\Bigr]y+\bar{\iota}'\,v_{{\bar{\alpha}\bar{\nu}}}\,z
\Bigr]\right)^{2\Delta_{{\bar{\alpha}\bar{\nu}}}^{\bar{\iota}'}-1} \, .
\label{C-v-1iotaD0}
\end{eqnarray}
\item When both $2\Delta_{\bar{\alpha}\bar{\nu}}^{\pm 1}=0$ and the remaining two parameters are
finite,
\begin{eqnarray}
& & F_0 (z) = D_0\,\sqrt{{v_{{\bar{\alpha}\bar{\nu}}}\over
v_{\alpha\nu}}}
\Theta\Bigl({1\over v_{\alpha\nu}}-\vert z\vert\Bigr)
\nonumber \\
& \times & \prod_{\iota' =\pm 1}{1\over \Gamma (2\Delta_{\alpha\nu}^{\iota'})} 
\left(\sqrt{{v_{{\bar{\alpha}\bar{\nu}}}\over
v_{\alpha\nu}}}\,\Bigl[1+\iota' v_{\alpha\nu}\,z
\Bigr]\right)^{2\Delta_{\alpha\nu}^{\iota'}-1} \, .
\label{C-v-2D0}
\end{eqnarray}
\item When both $2\Delta_{\alpha\nu}^{\pm 1}=0$ and the remaining two parameters are
finite,
\begin{equation}
F_0 (z) = D_0\,\sqrt{{v_{\alpha\nu}\over v_{{\bar{\alpha}\bar{\nu}}}}}
\prod_{\bar{\iota}' =\pm 1}{\Theta\Bigl({1\over v_{{\bar{\alpha}\bar{\nu}}}}-\bar{\iota}' z\Bigr)\over 
\Gamma (2\Delta_{{\bar{\alpha}\bar{\nu}}}^{\bar{\iota}'})} 
\left(\sqrt{{v_{\alpha\nu}\over v_{{\bar{\alpha}\bar{\nu}}}}}\,
\Bigl[1-\bar{\iota}' v_{{\bar{\alpha}\bar{\nu}}}\,z
\Bigr]\right)^{2\Delta_{{\bar{\alpha}\bar{\nu}}}^{\bar{\iota}'}-1} = 0 \, .
\label{C-v-2iotaD0}
\end{equation}
\item When $2\Delta_{\alpha\nu}^{-\iota}=2\Delta_{\bar{\alpha}\bar{\nu}}^{-\bar{\iota}}=0$ and the remaining 
two parameters are
finite,
\begin{eqnarray}
& & F_0 (z) = {2D_0\over\sqrt{v_{\alpha\nu} v_{\bar{\alpha}\bar{\nu}}}}
\left({v_{\alpha\nu} - \iota\bar{\iota}\, v_{\bar{\alpha}\bar{\nu}}\over
v_{\alpha\nu} v_{\bar{\alpha}\bar{\nu}}}\right)^{1-\zeta_0}
\nonumber \\
& \times & {\Theta\Bigl(\bar{\iota} [z -{\iota\over v_{\alpha\nu}}]\Bigr)\over
\Gamma (2\Delta_{{\bar{\alpha}\bar{\nu}}}^{\bar{\iota}})}
{\Theta\Bigl({1\over v_{{\bar{\alpha}\bar{\nu}}}}-\bar{\iota} z\Bigr)
\over \Gamma (2\Delta_{\alpha\nu}^{\iota})} \nonumber \\
& \times &
\left(2\sqrt{{v_{\alpha\nu}\over v_{{\bar{\alpha}\bar{\nu}}}}}\,
\bar{\iota} [z -{\iota\over v_{\alpha\nu}}]\right)^{2\Delta_{{\bar{\alpha}\bar{\nu}}}^{\bar{\iota}}-1}
\left(2\sqrt{{v_{{\bar{\alpha}\bar{\nu}}}\over
v_{\alpha\nu}}}\,[{1\over v_{{\bar{\alpha}\bar{\nu}}}}-\bar{\iota} z]
\right)^{2\Delta_{\alpha\nu}^{\iota}-1}  \, .
\label{C-v-1-1D0}
\end{eqnarray}
\item When both $2\Delta_{\alpha\nu}^{\iota}>0$ (and $2\Delta_{\bar{\alpha}\bar{\nu}}^{\bar{\iota}}>0$)
and the remaining three parameters vanish,
\begin{equation}
F_0 (z) = {D_0\over v_{\alpha\nu}\Gamma (2\Delta_{\alpha\nu}^{\iota})}
\left(2\sqrt{{v_{{\bar{\alpha}\bar{\nu}}}\over
v_{\alpha\nu}}}\right)^{2\Delta_{\alpha\nu}^{\iota}-1} 
\delta \left(z -{\iota\over v_{\alpha\nu}}\right) \, ,
\label{C-v-3D0}
\end{equation}
(and a similar expression with $\alpha\nu$,$\iota$, and 
$\bar{\alpha}\bar{\nu}$ replaced by $\bar{\alpha}\bar{\nu}$,
$\bar{\iota}$, and $\alpha\nu$, respectively).
\end{enumerate}



\begin{thebibliography}{99}
\bibitem{note}        
	While the first expression provided in Eq. (56) of Ref. \cite{V-1}
	is correct, there is a misprint in its second expression. It must 
	be multiplied by $2$ what gives our expressions of
	Eq. (\ref{B-J-i-sum-GG}). Such a misprint became an error that
	propagated to expression (62) of that reference which 
	must be multiplied by $4$. That gives our expression (\ref{C-v}) 
	of Appendix A for 
	$v_{c0}>v_{s1}$. Furthermore, it propagated to Ref. \cite{LE} so that the pre-factor 
	given in Eq. (50) of the latter reference must also be multiplied by $4$.
	Finally, while the exponent of expression (73) of Ref. \cite{V-1}
	is correct, the prefactor is a poor approximation.
\end{thebibliography}
\end{document}